\documentclass[11pt]{article}
\usepackage[dvips]{graphicx}
\usepackage{bm}
\usepackage{amsmath}
\usepackage{amssymb}
\usepackage{float}
\usepackage{url}
\renewcommand{\baselinestretch}{1.1}
\renewcommand{\thepage}{}
\makeatletter
\@addtoreset{equation}{section}
\renewcommand{\theequation}{\thesection.\@arabic\c@equation}
\makeatother
\renewcommand{\thefootnote}{\fnsymbol{footnote}}
\begin{document}
\begin{titlepage}
\title{
\vspace*{-4ex}
\hfill{}\\
\hfill
\begin{minipage}{3.5cm}
\end{minipage}\\
\bf Numerical universal solutions in $a$-gauge\\
 in open string field theory
\vspace{3ex}
}

\author{
Isao~{\sc Kishimoto}\footnote{ikishimo@rs.socu.ac.jp}
\\
\vspace{0.5ex}\\
\\
{\it Center for Liberal Arts and Sciences, Sanyo-Onoda City University,}\\
{\it Daigakudori 1-1-1, Sanyo-Onoda Yamaguchi 756-0884, Japan}
\vspace{2ex}
}

\date{}
\maketitle

%
\vspace{7ex}

\begin{abstract}
\normalsize
In bosonic open string field theory, we construct numerical universal solutions in $a$-gauge corresponding to ``double brane'' and ``ghost brane'' solutions in Siegel gauge in addition to the tachyon vacuum solution, and evaluate their gauge invariants, which are energy and gauge-invariant observable. The $a$-gauge condition, which contains a real parameter $a$, was introduced by Asano and Kato. In earlier works it has been applied to find the tachyon vacuum solution with the level truncation method up to level $14$. The ``double brane'' and ``ghost brane'' solutions were constructed by Kudrna and Schnabl in Siegel gauge, which corresponds to ($a=0$)-gauge, up to level $28$.
Starting from these solutions, by varying $a$ little by little, we have constructed numerical solutions in $a$-gauge for various values of $a$ including $a=\infty$ up to level $20$.
Contrary to naive expectation, the gauge invariants of ``double brane'' and ``ghost brane'' solutions in $a$-gauge seem to be non-constant for $a$.
In particular, although the normalized energy $E$ of the ``double brane'' solution in $a$-gauge is approximately two for $a\sim 0$, we find that $E$ becomes almost one for $0.5<a<1$.
The gauge-invariant observable also behaves similarly.
It might imply that the ``double brane'' solution varies to a single brane solution in such $a$-gauges.

\end{abstract}

\vspace{5ex}

\end{titlepage}

\renewcommand{\thepage}{\arabic{page}}
\renewcommand{\thefootnote}{\arabic{footnote}}
\setcounter{page}{1}
\setcounter{footnote}{0}
%
\renewcommand{\baselinestretch}{1}
\tableofcontents
\renewcommand{\baselinestretch}{1.1}

\section{Introduction and summary
\label{sec:Introduction}}

In bosonic open string field theory, where the action is given by
\begin{align}
S[\Psi]&=-\dfrac{1}{g^{2}}\left(\frac{1}{2}\langle\Psi,Q_{\rm B}\Psi\rangle
+\frac{1}{3}\langle\Psi,\Psi\ast\Psi\rangle\right),
\label{eq:action}
\end{align}
various solutions to the equation of motion,
\begin{align}
Q_{\rm B}\Psi+\Psi\ast\Psi=0,
\label{eq:EOM}
\end{align}
have been constructed after the discovery of Schnabl's analytic solution for tachyon condensation \cite{Schnabl:2005gv}.
Before that, the numerical tachyon vacuum solution in Siegel, $b_{0}|\Psi\rangle=0$, was found by Sen and Zwiebach \cite{Sen:1999nx} with the level truncation method.\footnote{See also Refs.~\cite{Gaiotto:2002wy,Kudrna:2019xnw}.}
Kudrna and Schnabl in Ref.~\cite{Kudrna:2018mxa} numerically constructed ``double brane'' and  ``ghost brane'' solutions in the twist-even SU(1,1)-singlet state space, which is in Siegel gauge, in addition to the tachyon vacuum solution. 
In a previous work \cite{Kishimoto:2020vfg} we generalized these solutions to those in theory around the Takahashi--Tanimoto identity-based solution (TT solution). Roughly, ``double brane'' and  ``ghost brane'' solutions seem to be consistent with the usual interpretation of the TT solution, although the numerical behavior is rather vaguer than the tachyon vacuum and perturbative vacuum solutions \cite{Takahashi:2003ppa,Kishimoto:2009nd,Kishimoto:2009hc}.
We should note that the abovementioned numerical solutions are all in Siegel gauge, and it is preferable to check these numerical solutions in other gauges in order to confirm their physical interpretation.\footnote{
We note that numerical solutions in Schnabl gauge, ${\cal B}_{0}|\Psi\rangle=0$, were investigated in Refs.~\cite{Arroyo:2017iis,AldoArroyo:2019hvj}.
}

Asano and Kato defined the $a$-gauge: \cite{Asano:2006hk}
\begin{align}
\left(-2J_+b_0+a{\tilde Q}b_0c_0\right)|\Psi\rangle=0
\label{eq:a-gauge_condition}
\end{align}
with a real parameter $a$ as a consistent gauge-fixing condition. Here, $J_+$ and ${\tilde Q}$ are given from the expansion of the Becchi--Rouet--Stora--Tyutin (BRST) operator with respect to the ghost zero modes:
\begin{align}
Q_{\rm B}=c_0L_0-2J_+b_0+{\tilde Q}.
\end{align}
In the case $a=0$, one can show that Eq.~(\ref{eq:a-gauge_condition}) is reduced to the Siegel gauge condition $b_0|\Psi\rangle=0$.
On the other hand, in the case $a=\infty$, Eq.~(\ref{eq:a-gauge_condition}) becomes ${\tilde Q}b_0c_0|\Psi\rangle=0$,
which corresponds to the Landau gauge in the massless sector.

Under the $a$-gauge condition, the tachyon vacuum solution has been investigated in earlier works \cite{Asano:2006hm,Kishimoto:2009cz,Kishimoto:2009hb} up to level $14$.\footnote{
The level is given by the eigenvalue of $L_0+1$.
}
We note that the condition in Eq.~(\ref{eq:a-gauge_condition}) does not mix the level of $\Psi$, and hence it is suitable for the level truncation method.
Furthermore, we can restrict the space of string fields to the universal space, which is spanned by the states made of  $(b,c)$-ghost modes and matter Virasoro generators on the conformal vacuum, for constructing solutions in $a$-gauge because Eq.~(\ref{eq:a-gauge_condition}) is compatible with it.
We also impose the twist-even condition, which implies the state space is spanned by even-level states.
In Siegel gauge, the solutions for tachyon vacuum, ``double brane,'' and ``ghost brane'' were constructed in the twist-even universal space. Hence, we can expect that the corresponding numerical solutions in $a$-gauge are in it. We notice that we cannot impose the SU(1,1)-singlet condition in $a$-gauge (for $a\ne 0$) due to incompatibility.

In this paper we explore three numerical solutions in $a$-gauge in the twist-even universal space obtained from tachyon vacuum, ``double brane,'' and ``ghost brane'' solutions in Siegel gauge up to level $20$.
We evaluate two gauge invariants, which are the energy $E[\Psi]$ and the gauge-invariant observable $E_{0}[\Psi]$ for the obtained solutions:
\begin{align}
&E[\Psi]=1-2\pi^2g^{2}S[\Psi],
&&E_0[\Psi]=1-2\pi \langle I|V|\Psi\rangle,
\label{eq:E-E0def}
\end{align}
where $|I\rangle$ is the identity string field and $V$ is a vertex operator $c\tilde{c}V^{\rm mat}(z,\bar{z})$ for the on-shell closed string state inserted at the midpoint of open string.\footnote{
$E_{0}[\Psi]$ is also called the Ellwood invariant \cite{Ellwood:2008jh} or gauge-invariant overlap \cite{Kawano:2008ry} in the literature.
$E_{0}[\Psi]$ is expected to coincide with $E[\Psi]$ for a class of solutions to the equation of motion $\Psi$ \cite{Baba:2012cs}.
}
They are normalized as
$E=E_{0}=0$ for the tachyon vacuum, which corresponds to no brane, and $E=E_{0}=1$ for the perturbative vacuum, which is the trivial solution $\Psi=0$ and corresponds to a single brane.

\begin{figure}[htbp]
\begin{center}
\includegraphics[width=16cm]{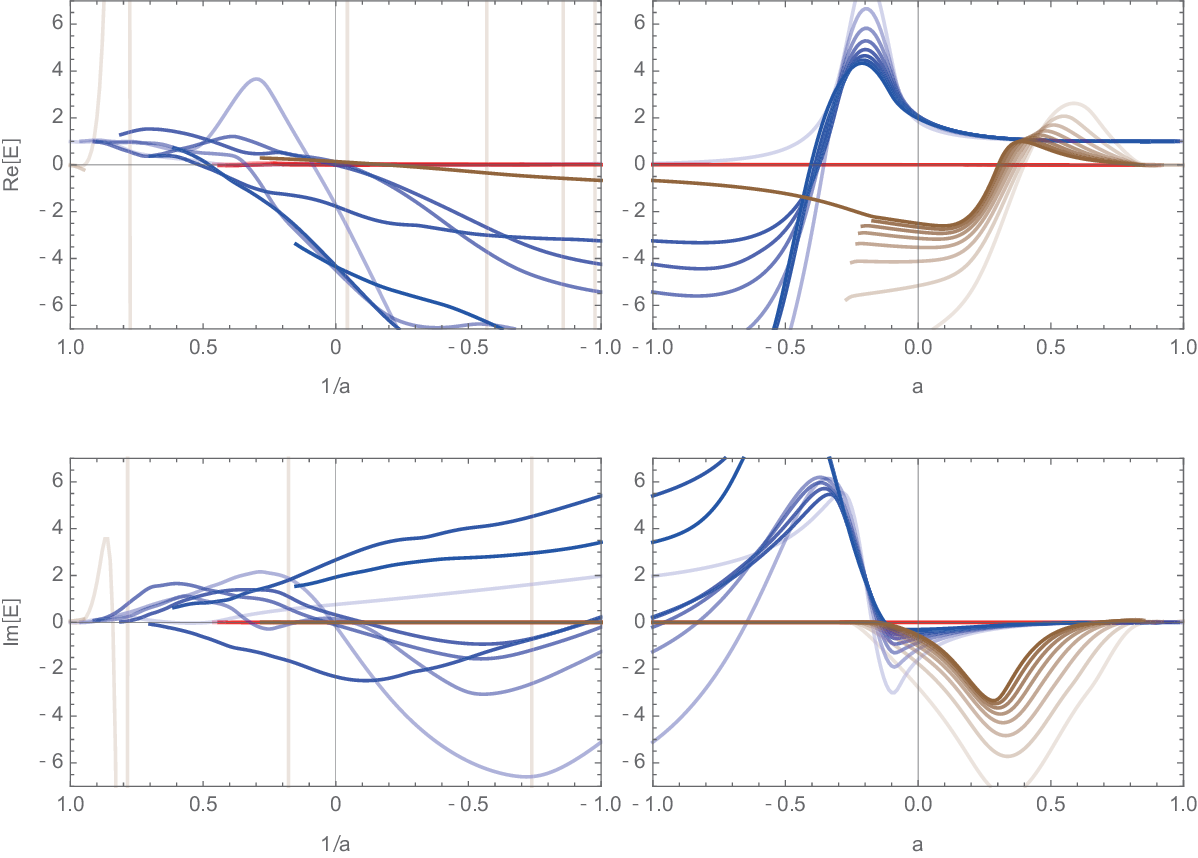}
\caption{Plots of the energy $E$, Eq.~(\ref{eq:E-E0def}), of solutions for various $a$ truncated up to level $L=6,8,\dots,20$, which are constructed from the tachyon vacuum solution (red), the ``double brane'' solution (blue), and the ``ghost brane'' solution (brown) in Siegel gauge ($a=0$). 
The upper and lower figures show the real and imaginary parts of $E$, respectively.
A darker color corresponds to a higher truncation level.
The imaginary part of $E$ for the tachyon vacuum (red) is exactly zero, thanks to the reality of the solution.
Plots for each solution are given in Figs.~\ref{fig:ReET}, \ref{fig:ED}, and \ref{fig:EG}.
\label{fig:ETGD}}
\end{center}
\end{figure}
\begin{figure}[htbp]
\begin{center}
\includegraphics[width=16cm]{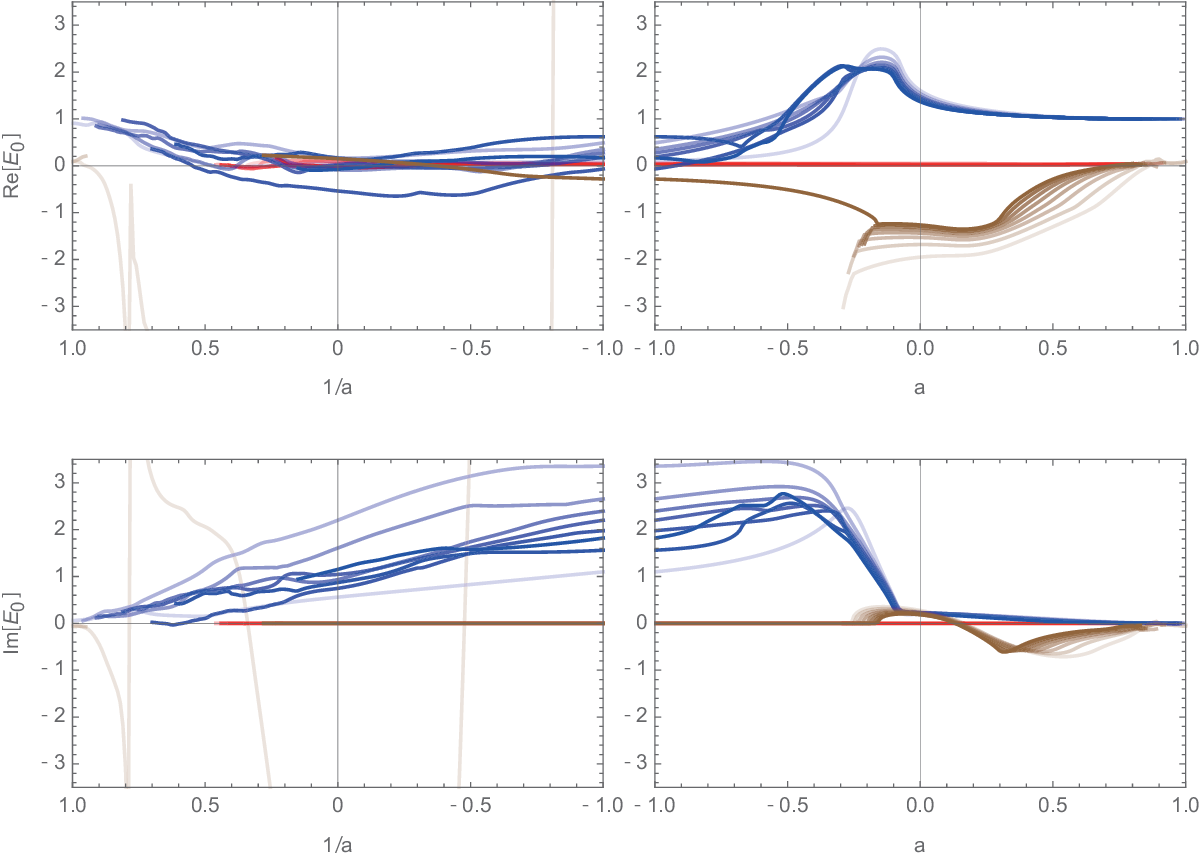}
\caption{Plots of the gauge-invariant observable $E_{0}$, Eq.~(\ref{eq:E-E0def}), of solutions for various $a$ truncated up to level $L=6,8,\dots,20$, which are constructed from the tachyon vacuum solution (red), the ``double brane'' solution (blue), and the ``ghost brane'' solution (brown) in Siegel gauge ($a=0$). The upper (lower) figures show the real (imaginary) part of $E_{0}$.
A darker color corresponds to a higher truncation level.
The imaginary part of $E_{0}$ for the tachyon vacuum (red) is exactly zero in the same way as $E$.
Note that the ranges of the vertical direction in the above plots are half of those in Fig.~\ref{fig:ETGD}.
Plots for each solution are given in Figs.~\ref{fig:ReE0T}, \ref{fig:E0D}, and \ref{fig:E0G}.
\label{fig:E0TGD}}
\end{center}
\end{figure}

In Figs.~\ref{fig:ETGD} and \ref{fig:E0TGD} we demonstrate resulting evaluations of Eq.~(\ref{eq:E-E0def}) for the above three solutions. In general, $E$ and $E_{0}$ become complex because ``double brane'' and ``ghost brane'' solutions in Siegel gauge do not satisfy the reality condition.
We confirmed that the tachyon vacuum exists for various values of $a$
because we found that the numerical solutions, which are constructed from the tachyon vacuum solution in Siegel gauge and satisfy the reality condition, give $E\simeq E_{0}\simeq 0$ as in Figs.~\ref{fig:ETGD} and \ref{fig:E0TGD}.
In the case of the solutions constructed from the ``double brane'' solution in Siegel gauge, $E\sim 2$ and $E_{0}\sim 1.4$ for $a\sim 0$, but $E\simeq E_{0}\simeq 1$ for $0.5<a<1$. This might imply that the ``double brane'' solution seems to vary to a ``single brane'' solution by varying the value of $a$.
In principle, $E[\Psi]$ and $E_{0}[\Psi]$ in Eq.~(\ref{eq:E-E0def}) are gauge invariant and therefore they should be constant with respect to the parameter $a$ if solutions are physically equivalent. Hence, the solutions might be in different branches for different values of $a$. In this sense, we could not confirm that the ``double brane'' solution indicates a double brane in the literal meaning. As for the solutions constructed from the ``ghost brane'' solution in Siegel gauge, $E\sim -2.5$ and $E_{0}\sim -1.3$ for $a\simeq 0$, and $E$ and $E_{0}$ vary to $0$ for $a\sim 0.8$. These solutions also might be in different branches for different values of $a$ if they are valid numerical solutions. 

This paper is organized as follows.
In Sect.~\ref{sec:Newton} we explain our procedure for constructing numerical solutions in $a$-gauge. In Sect.~\ref{sec:results}, we show plots of numerical results for three solutions in various $a$-gauges. In Sect.~\ref{sec:remarks} we present some concluding remarks on our numerical results. Some details about our numerical calculations are shown in Appendix \ref{sec:App}.

\section{Construction of numerical solutions in $a$-gauge
 \label{sec:Newton}}
 
Here, we explain our procedure for constructing numerical solutions in $a$-gauge with the level truncation method.
We consider the twist-even universal space for numerical computation as mentioned in Sect.~\ref{sec:Introduction},
which is spanned by
\begin{align}
&L_{-l_1}^{\rm mat}\cdots L_{-l_m}^{\rm mat}
b_{-n_1}\cdots b_{-n_k}c_{-m_1}\cdots c_{-m_k}c_1|0\rangle,
\label{eq:spanned_by}
\\
&l_1\ge \cdots \ge l_m\ge 2,\qquad
n_1> \cdots > n_k\ge 1,\qquad
m_1>\cdots>m_k\ge 0,\\
&\sum_{j=1}^ml_j+\sum_{j=1}^kn_j+\sum_{j=1}^km_j=0,2,\dots, L,
\end{align}
where $L$ is the truncation level and is an even integer.
In particular, we take a normalized basis $(\psi_i,\chi_b)$, where $\psi_i$ does not contain $c_0$ and $\chi_b$ contains $c_0$, and they are multiplied by \sloppy\\ 
$\left|{\rm bpz}(L_{-l_1}^{\rm mat}\cdots L_{-l_m}^{\rm mat}|0\rangle_{\rm mat})L_{-l_1}^{\rm m}\cdots L_{-l_m}^{\rm mat}|0\rangle_{\rm m}\right|^{-1/2}$ with ${}_{\rm m}\langle 0|0\rangle_{\rm m}=1$
for the above form.
Using this basis, a string field $\Psi$ with ghost number $1$ in the twist-even universal space truncated up to level $L$ is expanded as
\begin{align}
\Psi&=\sum_{l=0}^{L/2}\sum_{i_{l}=1}^{D_{1,2l}}t_{i_{l}}\psi_{i_{l}}
+\sum_{l=0}^{L/2}\sum_{b_{l}=1}^{D_{2,2l}}u_{b_{l}}\chi_{b_{l}},
\label{eq:Phi_L1exp}
\end{align}
where $\psi_{i_{l}}$ and $\chi_{b_{l}}$ are states with level $2l$.
$D_{1,2l}$ and $D_{2,2l}$ are given in Table~\ref{tab:d1Ld2L} in Appendix \ref{sec:dimension}.
The coefficients $t_{i_{l}}$ and $u_{b_{l}}$ are complex numbers in general,
although they should be real if we impose the reality condition for $\Psi$.

\subsection{($a\ne \infty$)-gauge
\label{sec:aneinfty}}

In the case $a\ne\infty$,
imposing the $a$-gauge condition in Eq.~(\ref{eq:a-gauge_condition}) on Eq.~(\ref{eq:Phi_L1exp}), we have the following relation among the coefficients:
\begin{align}
&-2A{\bm u}+a{\tilde A}{\bm t}={\bm 0},
&&
{\bm t}=\begin{pmatrix}
t_{i_{1}}\\
t_{i_{2}}\\
\vdots\\
t_{i_{D_{1,2l}}}
\end{pmatrix},\qquad
{\bm u}=\begin{pmatrix}
u_{b_{1}}\\
u_{b_{2}}\\
\vdots\\
u_{b_{D_{2,2l}}}
\end{pmatrix},
\label{eq:relation-ut}
\end{align}
for each level $2l$, where $A$ is a $D_{2,2l}\times D_{2,2l}$ real symmetric matrix and $\tilde{A}$ is a $D_{2,2l}\times D_{1,2l}$ real matrix.
They are given by the Belavin--Polyakov--Zamolodchikov (BPZ) inner products
\begin{align}
&\langle\chi_{b_{l}},J_+b_0\chi_{c_{l}}\rangle=[A]_{b_{l}c_{l}},
&&\langle\chi_{b_{l}},\tilde{Q}b_0c_0\psi_{i_{l}}\rangle=[\tilde A]_{b_{l}i_{l}}.
\label{eq:AtilAmat}
\end{align}
Using the relation in Eq.~(\ref{eq:relation-ut}), we have an expression for a string field $\Psi_{a}$ in $a$-gauge:
\begin{align}
\Psi_{a}&=\sum_{l=0}^{L/2}\sum_{i_{l}=1}^{D_{1,2l}}\psi_{i_{l}}t_{i_{l}}
+\dfrac{a}{2}\sum_{l=0}^{L/2}\sum_{b_{l}=1}^{D_{2,2l}}\chi_{b_{l}}[A^{-1}{\tilde A}\,{\bm t}]_{b_{l}}.
\label{eq:Psi_adef}
\end{align}
Substituting the above into the action in Eq.~(\ref{eq:action}),
we have an equation of motion which is obtained from a partial derivative of $S[\Psi_{a}]$ with respect to $t_{i_{l}}$:
\begin{align}
\langle\psi_{i_l}
+\dfrac{a}{2}\sum_{b_l=1}^{D_{2,2l}}\chi_{b_{l}}[A^{-1}\tilde{A}]_{b_li_l},
Q_{\rm B}\Psi_{a}+\Psi_{a}\ast\Psi_{a}\rangle=0.
\label{eq:EOM_AK}
\end{align}
This gives a system of $N_{1,L}$ equations with $N_{1,L}$ variables $t_{i_{l}}$, where $N_{1,L}$ is given in Table~\ref{tab:univedim} in Appendix \ref{sec:dimension}.
We solve the equation using Newton's method.
Linearizing Eq.~(\ref{eq:EOM_AK}), we have an equation for $\Psi_{a}^{(n+1)}$:
\begin{align}
&\langle\psi_{i_l}+\dfrac{a}{2}\sum_{b_l=1}^{D_{2,2l}}\chi_{b_l}[A^{-1}\tilde{A}]_{b_li_l},
Q_{\rm B}\Psi_{a}^{(n+1)}+\Psi_{a}^{(n+1)}\ast\Psi_{a}^{(n)}+\Psi_{a}^{(n)}\ast\Psi_{a}^{(n+1)}\rangle\nonumber\\
&\qquad =
\langle\psi_{i_l}+\dfrac{a}{2}\sum_{b_l=1}^{D_{2,2l}}\chi_{b_l}[A^{-1}\tilde{A}]_{b_li_l},
\Psi_{a}^{(n)}\ast\Psi_{a}^{(n)}\rangle
\label{eq:EOM_AKite}
\end{align}
with known $\Psi_{a}^{(n)}$,
and we solve it iteratively
with an appropriate initial configuration $\Psi_{a}^{(0)}$.
If it converges to a configuration in the large-$n$ limit,
$\Psi_{a}^{(\infty)}$ gives a solution to Eq.~(\ref{eq:EOM_AK}).

\subsection{($a\ne 0$)-gauge}
 
 In order to treat the case of large $a$-gauge numerically, the procedure in Sect.~\ref{sec:aneinfty} is not appropriate.
 Here, we solve the $a$-gauge condition in Eq.~(\ref{eq:a-gauge_condition}) around $a=\infty$ by using the singular-value decomposition of $\tilde{A}$ defined in Eq.~(\ref{eq:AtilAmat}):
 \begin{align}
 &\tilde{A}=U\hat{\Lambda}V^{\rm T},
\qquad\hat{\Lambda}=(\Lambda,O),\\
&[\Lambda]_{b_{l}c_{l}}=\lambda_{b_{l}}\delta_{b_{l},c_{l}},
\qquad \lambda_1\ge \lambda_2\ge \cdots\ge \lambda_{D_{2,2l}}>0,
 \end{align}
 for each level $2l$, where $U$ is a $D_{2,2l}\times D_{2,2l}$ real orthogonal matrix and $V$ is a $D_{1,2l}\times D_{1,2l}$ real orthogonal matrix. $O$ in $\hat{\Lambda}$ is $D_{2,2l}\times (D_{1,2l}-D_{2,2l})$ zero matrix.
Using these, we define a new basis $(\hat\psi_{i_{l}},\hat\chi_{b_{l}})$: 
\begin{align}
&\hat{\psi}_{i_{l}}=\sum_{j_{l}=1}^{D_{1,2l}}\psi_{j_{l}}[V]_{j_{l}i_{l}},\qquad\qquad
\hat{\chi}_{b_{l}}=\sum_{c_{l}=1}^{D_{2,2l}}\chi_{c_{l}}[U]_{c_{l}b_{l}},
\end{align}
and a string field of the form in Eq.~(\ref{eq:Phi_L1exp}) is rewritten as
\begin{align}
&\Psi=\sum_{l=0}^{L/2}\sum_{i_{l}=1}^{D_{1,2l}}\hat{t}_{i_{l}}\hat{\psi}_{i_{l}}
+\sum_{l=0}^{L/2}\sum_{b_{l}=1}^{D_{2,2l}}\hat{u}_{b_{l}}\hat{\chi}_{b_{l}},
&&\hat{t}_{i_{l}}=\sum_{j_{l}=1}^{D_{1,2l}}t_{j_{l}}[V]_{j_{l}i_{l}},\quad\hat{u}_{b_{l}}=\sum_{c_{l}=1}^{D_{2,2l}}u_{c_{l}}[U]_{c_{l}b_{l}}.
\label{eq:Phi_L1exphat}
\end{align}
Imposing the $a$-gauge condition from Eq.~(\ref{eq:a-gauge_condition}) on the above, we have the following relation among the coefficients:
\begin{align}
&-2\hat{A}\hat{\bm u}+a\hat{\Lambda}\hat{\bm t}={\bm 0},
&&
\hat{\bm t}=\begin{pmatrix}
\hat{t}_{i_{1}}\\
\hat{t}_{i_{2}}\\
\vdots\\
\hat{t}_{i_{D_{1,2l}}}
\end{pmatrix},\qquad
\hat{\bm u}=\begin{pmatrix}
\hat{u}_{b_{1}}\\
\hat{u}_{b_{2}}\\
\vdots\\
\hat{u}_{b_{D_{2,2l}}}
\end{pmatrix}
\label{eq:relation-uthat}
\end{align}
for each level $2l$, where $\hat{A}=U^{\rm T}AU$ is a $D_{2,2l}\times D_{2,2l}$ real symmetric matrix.
In the case $a\ne 0$,
we decompose $\hat{\bm t}$ as
\begin{align}
&\hat{\bm t}=
\begin{pmatrix}
{\bm w}\\
{\bm v}
\end{pmatrix},
\qquad
{\bm w}=
\begin{pmatrix}
w_1\\
w_2\\
\vdots\\
w_{D_{2,L}}
\end{pmatrix},
\qquad
{\bm v}=
\begin{pmatrix}
v_1\\
v_2\\
\vdots\\
v_{D_{1,2l}-D_{2,2l}}
\end{pmatrix},
\label{eq:wvvecdef}
\end{align}
and the relation (\ref{eq:relation-uthat}) can be rewritten as
\begin{align}
{\bm w}=2\hat{a}\Lambda^{-1}\hat{A}\hat{\bm u}=2\hat{a}\Lambda^{-1}U^{\rm T}A{\bm u},
\end{align}
with $\hat{a}=1/a$. There is no constraint on $v_{i_{l}^{\prime}}$ ($i_{l}^{\prime}=1,2,\dots,D_{1,2l}-D_{2,2l}$). From the above, for $a\ne 0$ we have an expression for a string field $\varPsi_{\hat{a}}$ in $a$-gauge:
\begin{align}
&\varPsi_{\hat a}
=\sum_{l=0}^{L/2}\sum_{i^{\prime}_{l}=1}^{D_{1,2l}-D_{2,2l}}\varphi_{i^{\prime}_{l}}v_{i^{\prime}_{l}}
+\sum_{l=0}^{L/2}\sum_{b_{l}=1}^{D_{2,2l}}
\left(\chi_{b_{l}}
+2{\hat a}[{\bm \phi}\Lambda^{-1}U^{\rm T}A]_{b_{l}}\right)u_{b_{l}},
\qquad\qquad \hat{a}=\frac{1}{a},
\label{eq:varPsiahat}
\\
&\hat{\bm \psi}={\bm \psi}V=({\bm \phi},{\bm \varphi}),
\qquad{\bm \phi}=(\hat{\psi}_1,\hat{\psi}_2,\cdots,\hat{\psi}_{D_{2,2l}}),
\qquad{\bm \varphi}=(\hat{\psi}_{D_{2,2l}+1},\hat{\psi}_{D_{2,2l}+2},\cdots,\hat{\psi}_{D_{1,2l}}).
\end{align}

In the case that $a\ne 0$ and $a\ne \infty$, we have the following relations among the coefficients of Eqs.~(\ref{eq:Psi_adef}) and (\ref{eq:varPsiahat}):
\begin{align}
&{\bm t}=V
\begin{pmatrix}
\dfrac{2}{a}\Lambda^{-1}U^{\rm T}A{\bm u}\\
{\bm v}
\end{pmatrix}
\qquad\Leftrightarrow
\qquad
{\bm v}=(O,I)V^{\rm T}
{\bm t},\quad
{\bm u}=\dfrac{a}{2}A^{-1}\tilde{A}{\bm t},
\label{eq:t2uv}
\end{align}
by identifying $\Psi_{a}$ with $\varPsi_{\hat{a}=1/a}$.
In the above ${\bm v}$, $O$ is the $(D_{1,2l}-D_{2,2l})\times D_{2,2l}$ zero matrix and $I$ is the identity matrix of size $(D_{1,2l}-D_{2,2l})$.

Substituting Eq.~(\ref{eq:varPsiahat}) into the action in Eq.~(\ref{eq:action}), we have equations of motion which are obtained from partial derivatives of $S[\varPsi_{\hat a}]$ with respect to $v_{i^{\prime}_{l}}$ and $u_{b_{l}}$:
\begin{align}
&\langle \varphi_{i_l^{\prime}},Q_{\rm B}\varPsi_{\hat{a}}+\varPsi_{\hat{a}}\ast\varPsi_{\hat{a}}
\rangle=0,
\label{eq:EOMhata1}\\
&\langle \chi_{b_l}+2\hat{a}\sum_{c_l=1}^{D_{2,2l}}\phi_{c_l}[\Lambda^{-1}U^{\rm T}A]_{c_lb_l}
,Q_{\rm B}\varPsi_{\hat{a}}+\varPsi_{\hat{a}}\ast\varPsi_{\hat{a}}
\rangle=0.
\label{eq:EOMhata2}
\end{align}
These give a system of $(N_{1,L}-N_{2,L})+N_{2,L}=N_{1,L}$ equations with $N_{1,L}$ variables $(v_{i^{\prime}_{l}},u_{b_{l}})$, where $N_{g,L}$ is defined in Eq.~(\ref{eq:NgLdef}).
Linearizing them, we have equations for $\varPsi^{(n+1)}_{\hat a}$ for Newton's method:
\begin{align}
&\langle \varphi_{i_l^{\prime}},Q_{\rm B}\varPsi_{\hat{a}}^{(n+1)}
+\varPsi_{\hat{a}}^{(n+1)}\ast\varPsi_{\hat{a}}^{(n)}
+\varPsi_{\hat{a}}^{(n)}\ast\varPsi_{\hat{a}}^{(n+1)}
\rangle=
\langle \varphi_{i_l^{\prime}},
\varPsi_{\hat{a}}^{(n)}\ast\varPsi_{\hat{a}}^{(n)}\rangle,
\label{eq:EOMhata1ite}
\\
&\langle \chi_{b_l}+2\hat{a}\sum_{c_l=1}^{D_{2,2l}}\phi_{c_l}[\Lambda^{-1}U^{\rm T}A]_{c_lb_l}
,Q_{\rm B}\varPsi_{\hat{a}}^{(n+1)}
+\varPsi_{\hat{a}}^{(n+1)}\ast\varPsi_{\hat{a}}^{(n)}
+\varPsi_{\hat{a}}^{(n)}\ast\varPsi_{\hat{a}}^{(n+1)}
\rangle\nonumber\\
&\qquad 
=\langle \chi_{b_l}+2\hat{a}\sum_{c_l=1}^{D_{2,2l}}\phi_{c_l}[\Lambda^{-1}U^{\rm T}A]_{c_lb_l},
\varPsi_{\hat{a}}^{(n)}\ast\varPsi_{\hat{a}}^{(n)}
\rangle;
\label{eq:EOMhata2ite}
\end{align}
these have known $\varPsi^{(n)}_{\hat a}$, and we solve them iteratively with an appropriate initial configuration $\varPsi^{(0)}_{\hat a}$. If it converges to a configuration in the large-$n$ limit, $\varPsi^{(\infty)}_{\hat a}$ gives a solution to Eqs.~(\ref{eq:EOMhata1}) and (\ref{eq:EOMhata2}).

\subsection{Validity of solutions}

In general, the equation of motion to the action in Eq.~(\ref{eq:action}) is given by Eq.~(\ref{eq:EOM}).
We should note that Eq.~(\ref{eq:EOM_AK}) for $a\ne \infty$ is only a part of the equation of motion in Eq.~(\ref{eq:EOM}). In the same sense, Eqs.~(\ref{eq:EOMhata1}) and (\ref{eq:EOMhata2}) for $a\ne 0$ are only a part of the equation of motion in Eq.~(\ref{eq:EOM}).
Hence, we should confirm the remaining part of the equation of motion for numerical solutions to Eq.~(\ref{eq:EOM_AK}) for $a\ne \infty$ or Eqs.~(\ref{eq:EOMhata1}) and (\ref{eq:EOMhata2}) for $a\ne 0$. Here, we call such equations for the validity of numerical solutions in $a$-gauge the out-of-$a$-gauge equations.\footnote{In the case of the Siegel gauge, they were investigated for the tachyon vacuum solution in Ref.~\cite{Hata:2000bj} as the BRST invariance.
In Ref.~\cite{Kudrna:2018mxa}, they were called the out-of-Siegel-gauge equations, and their lowest level was denoted as $\Delta_{S}$ and evaluated for various solutions in Siegel gauge.
}

In the case of solutions to Eq.~(\ref{eq:EOM_AK}) for $a\ne \infty$, the out-of-$a$-gauge equations are
\begin{align}
&\langle \chi_{b_l},Q_{\rm B}\Psi_{a}+\Psi_{a}\ast\Psi_{a}\rangle=0.
\label{eq:outofgauge}
\end{align}
They consist of $N_{2,L}$ equations.
In fact, Eqs.~(\ref{eq:EOM_AK}) and (\ref{eq:outofgauge}) imply the equation of motion in Eq.~(\ref{eq:EOM}): $Q_{\rm B}\Psi_{a}+\Psi_{a}\ast\Psi_{a}=0$.
In particular, the lowest level of $\chi_{b_l}$ is $2l=2$, and there is only one $\chi_{b_1}$ due to $D_{2,2}=1$.
We denote the left hand side of Eq.~(\ref{eq:outofgauge}) for $\chi_{b_1}$ as $\Delta_{a}[\Psi_{a}]$.

Similarly, in the case of solutions to Eqs.~(\ref{eq:EOMhata1}) and (\ref{eq:EOMhata2}) for $a\ne 0$, the out-of-$a$-gauge equations are
\begin{align}
&\langle\phi_{b_l},Q_{\rm B}\varPsi_{\hat{a}}+\varPsi_{\hat{a}}\ast\varPsi_{\hat{a}}\rangle=0.
\label{eq:outofgaugehat}
\end{align}
They consist of $N_{2,L}$ equations.
Equations (\ref{eq:EOMhata1}), (\ref{eq:EOMhata2}), and (\ref{eq:outofgaugehat}) imply the equation of motion in Eq.~(\ref{eq:EOM}): $Q_{\rm B}\varPsi_{\hat{a}}+\varPsi_{\hat{a}}\ast\varPsi_{\hat{a}}=0$.
The lowest level of $\phi_{b_l}$ is $2l=2$, and there is only one $\phi_{b_1}$.
We denote the left hand side of Eq.~(\ref{eq:outofgaugehat}) for $\phi_{b_1}$ as $\Delta_{\hat a}[\varPsi_{\hat a}]$.\\

The reality condition of the open string field is imposed for the reality of the action.
However, numerical solutions for the ``double brane'' and ``ghost brane'' in Siegel gauge are constructed from complex initial configurations, and then coefficients in Eqs.~(\ref{eq:Psi_adef}) and (\ref{eq:varPsiahat}) become complex in general.

In order to check the reality condition,
we define ${\rm Im}/{\rm Re}_{a}[\Psi_{a}]$ for Eq.~(\ref{eq:Psi_adef}) and
${\rm Im}/{\rm Re}_{\hat a}[\varPsi_{\hat a}]$ for Eq.~(\ref{eq:varPsiahat}) 
using the Euclidean norm:
\begin{align}
&{\rm Im}/{\rm Re}_{a}[\Psi_{a}]=\frac{\sqrt{\sum_{l=0}^{L/2}\sum_{i_{l}=1}^{D_{1,2l}}({\rm Im}(t_{i_{l}}))^{2}}}{\sqrt{\sum_{l=0}^{L/2}\sum_{i_{l}=1}^{D_{1,2l}}({\rm Re}(t_{i_{l}}))^{2}}}\,,
\label{eq:IoRadef}\\
&{\rm Im}/{\rm Re}_{\hat a}[\varPsi_{\hat a}]
=\frac{\sqrt{\sum_{l=0}^{L/2}\sum_{i^{\prime}_{l}=1}^{D_{1,2l}-D_{2,2l}}({\rm Im}(v_{i^{\prime}_{l}}))^{2}
+\sum_{l=0}^{L/2}\sum_{b_{l}=1}^{D_{2,2l}}({\rm Im}(u_{b_{l}}))^{2}}
}{\sqrt{\sum_{l=0}^{L/2}\sum_{i^{\prime}_{l}=1}^{D_{1,2l}-D_{2,2l}}({\rm Re}(v_{i^{\prime}_{l}}))^{2}
+\sum_{l=0}^{L/2}\sum_{b_{l}=1}^{D_{2,2l}}({\rm Re}(u_{b_{l}}))^{2}}
}\,.
\label{eq:IoRahatdef}
\end{align}
They should be zero if $\Psi_{a}$ and $\varPsi_{\hat a}$ satisfy the reality condition.

\section{Results of calculation
\label{sec:results}}

\subsection{Initial configurations
\label{sec:initialconfig}
}

We have constructed three solutions in various $a$-gauges starting from the tachyon vacuum solution $\Psi^{\rm T}_{a=0}$, the ``double brane'' solution $\Psi^{\rm D}_{a=0}$, and the ``ghost brane'' solution $\Psi^{\rm G}_{a=0}$ in Siegel gauge, namely in ($a=0$)-gauge.
The initial configurations for each procedure of Newton's method were chosen as follows.

\begin{itemize}
\item We construct numerical solutions at $a=0$, $\Psi^{\rm T}_{a=0}$, $\Psi^{\rm D}_{a=0}$, and $\Psi^{\rm G}_{a=0}$, truncated up to level $20$, using Newton's method level by level from the lowest.
$\Psi^{\rm T}_{a=0}$ is obtained from a unique nontrivial real solution at level $0$. 
$\Psi^{\rm D}_{a=0}$ is obtained from one of the complex solutions at the truncation level $2$. 
$\Psi^{\rm G}_{a=0}$ is obtained from one of the complex solutions at the truncation level $4$.

\item At each truncation level, from solutions for $a=0$, $\Psi^{\rm T}_{a=0}$, $\Psi^{\rm D}_{a=0}$, and $\Psi^{\rm G}_{a=0}$, we construct solutions to Eq.~(\ref{eq:EOM_AK}), $\Psi^{\rm T}_{a}$, $\Psi^{\rm D}_{a}$, and $\Psi^{\rm G}_{a}$, varying $a$ little by little up to $a=\pm 1$. We choose an initial configuration as $\Psi_{a+\varDelta a}^{(0)}=\Psi_{a}^{(\infty)}$, where we adopt $\varDelta a=\pm 0.01$ as a difference of $a$.

\item At each truncation level, after solutions for $a=\pm 1$ are obtained, we construct solutions to Eqs.~(\ref{eq:EOMhata1}) and (\ref{eq:EOMhata2}), $\varPsi^{\rm T}_{\hat a}$, $\varPsi^{\rm D}_{\hat a}$, and $\varPsi^{\rm G}_{\hat a}$, varying $\hat a$ little by little from solutions for $\hat{a}=\frac{1}{a}=\pm 1$, with the relations of coefficients in Eq.~(\ref{eq:t2uv}).
We choose an initial configuration as $\varPsi_{\hat{a}+\varDelta \hat{a}}^{(0)}=\varPsi_{\hat a}^{(\infty)}$, where we adopt $\varDelta\hat{a}=\mp 0.01$ as a difference of $\hat{a}$.

\end{itemize}

The above calculations were terminated if iterations of Newton's method did not converge for particular values of $a$ (or $\hat{a}$) at each truncation level.
In numerical computation, we stopped the iteration Eq.~(\ref{eq:EOM_AKite}) (or Eqs.~(\ref{eq:EOMhata1ite}) and (\ref{eq:EOMhata2ite}))
if $\|\Psi_{a}^{(n+1)}-\Psi_{a}^{(n)}\|/\|\Psi_{a}^{(n)}\| < \varepsilon$
(or $\|\varPsi_{\hat a}^{(n+1)}-\varPsi_{\hat a}^{(n)}\|/\|\varPsi_{\hat a}^{(n)}\| < \varepsilon$)
with the Euclidean norm of $(t_{i_{l}})$ (or $(v_{i^{\prime}_{l}},u_{b_{l}})$) and we adopted $\Psi_{a}^{(n+1)}$ (or $\varPsi_{\hat a}^{(n+1)}$)
as a solution to Eq.~(\ref{eq:EOM_AK}) (or Eqs.~(\ref{eq:EOMhata1}) and (\ref{eq:EOMhata2})): $\Psi_{a}^{(\infty)}$ (or $\varPsi_{\hat a}^{(\infty)}$).
We adopted $\varepsilon = 5.0\times 10^{-12}$ and the maximum number of iterations as $15$ with the long double format in our C++ code.
In the figures for plots of various quantities in this paper, the horizontal direction denotes the value of $a$ or ${\hat a}=1/a$, and adjacent data points for each truncation level are joined with line segments.

\subsection{Tachyon vacuum solution
\label{sec:Tsol}
}

\begin{figure}[htbp]
\begin{center}
\includegraphics[width=16cm]{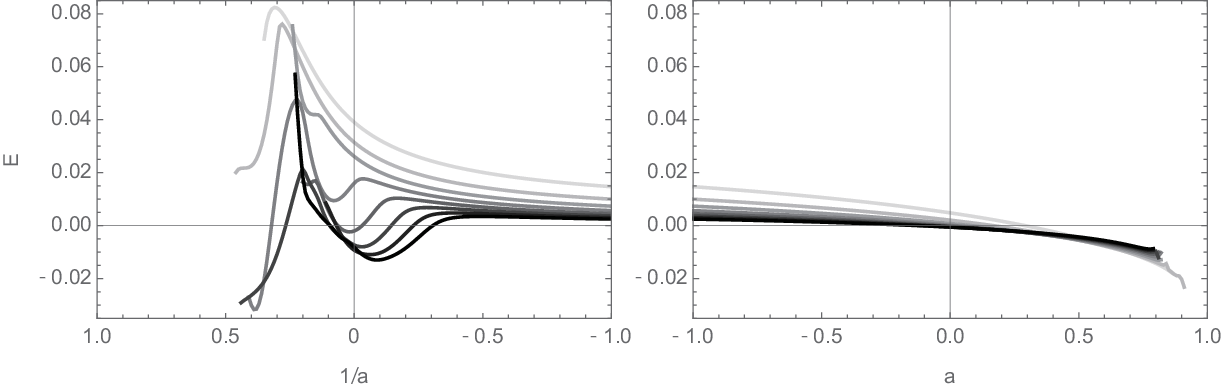}
\caption{Plots of the energy $E$ from Eq.~(\ref{eq:E-E0def}) of the tachyon vacuum solutions, $\varPsi_{{\hat a}=1/a}^{\rm T}$ (left) and $\Psi_{a}^{\rm T}$ (right), truncated up to level $L=6,8,\dots,20$.
A darker line corresponds to a higher level.
\label{fig:ReET}
}
\end{center}
\end{figure}

\begin{figure}[htbp]
\begin{center}
\includegraphics[width=16cm]{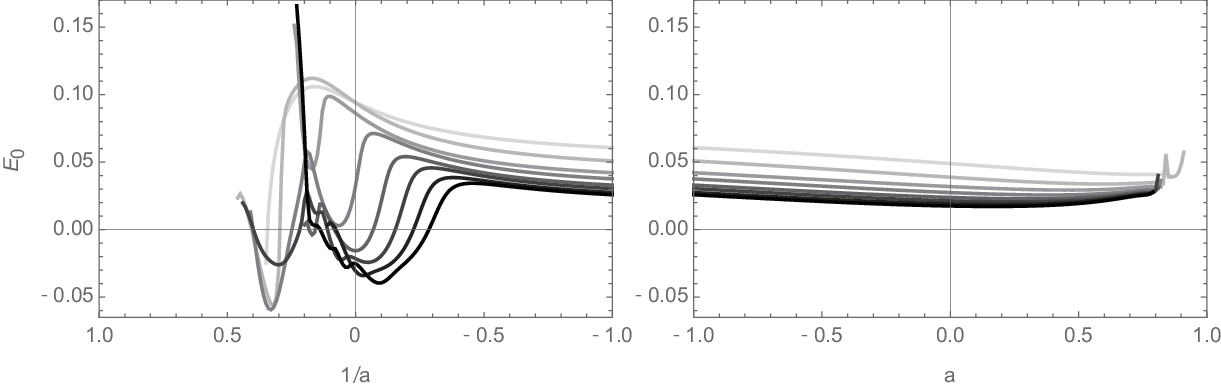}
\caption{Plots of the gauge-invariant observable $E_{0}$ from Eq.~(\ref{eq:E-E0def}) of the tachyon vacuum solutions, $\varPsi_{{\hat a}=1/a}^{\rm T}$ (left) and $\Psi_{a}^{\rm T}$ (right), truncated up to level $L=6,8,\dots,20$.
A darker line corresponds to a higher level.
\label{fig:ReE0T}
}
\end{center}
\end{figure}

The tachyon vacuum solution: $\Psi_{a}^{\rm T}$ (or $\varPsi_{\hat a}^{\rm T}$) in various $a$-gauges has been constructed from the conventional $\Psi_{a=0}^{\rm T}$ in the Siegel gauge as in Sect.~\ref{sec:initialconfig}. 
At the truncation level $L=20$, which is the highest one in our computation, $\Psi_{a}^{\rm T}$ was obtained for $-1\le a\le 0.79$, and $\varPsi_{\hat a}^{\rm T}$, which was constructed through the solution for $\hat{a}=1/a=-1$, was obtained for $-1\le \hat{a}\le 0.23$.\footnote{See Table~\ref{tab:a-range} in Appendix \ref{sec:detailednumericaldata} for lower levels.}
Plots of the gauge invariants $E$ and $E_{0}$ from Eq.~(\ref{eq:E-E0def}) of the solutions
$\varPsi_{\hat a}^{\rm T}$ and $\Psi_{a}^{\rm T}$ are given in Figs.~\ref{fig:ReET} and \ref{fig:ReE0T}. 
For various values of $a$, both $E$ and $E_{0}$ approach $0$ with increasing truncation level in a similar way to those of the tachyon vacuum solution in Siegel gauge, which are shown at $a=0$ in Figs.~\ref{fig:ReET} and \ref{fig:ReE0T}, respectively.

As in Fig.~\ref{fig:DeltaaT} in Appendix \ref{sec:validityagauge}, the lowest level of the out-of-$a$-gauge equations, $\Delta_{a}[\Psi^{\rm T}_{a}]$ or $\Delta_{\hat a}[\varPsi^{\rm T}_{\hat a}]$, approaches $0$ for $-1\le a\le 0.5$ or $-0.5\ge {\hat a}=1/a\ge -1$ respectively with increasing truncation level. This implies that $\Psi^{\rm T}_{a}$ and $\varPsi^{\rm T}_{\hat a}$ in these regions of $a$ are valid as solutions to the equation of motion in Eq.~(\ref{eq:EOM}).
We note that $\Psi^{\rm T}_{a}$ and $\varPsi^{\rm T}_{\hat a}$ satisfy the reality condition of the open string field from the beginning.

From the above, we expect that $\Psi_{a}^{\rm T}$ and $\varPsi_{\hat a}^{\rm T}$ represent the tachyon vacuum in $a$-gauge for various values of $a$, except $a\sim 1$ where we could not obtain solutions.
Here, we notice that at $a=1$ the $a$-gauge condition in Eq.~(\ref{eq:a-gauge_condition}) becomes $b_{0}c_{0}Q_{\rm B}|\Psi\rangle=0$, which is not suitable for a gauge-fixing condition in the free theory.
In this sense, the ($a=1$)-gauge might not be appropriate to construct numerical solutions.

\subsection{``Double brane'' solution
\label{sec:Dsol}
}

\begin{figure}[h]
\begin{center}
\includegraphics[width=16cm]{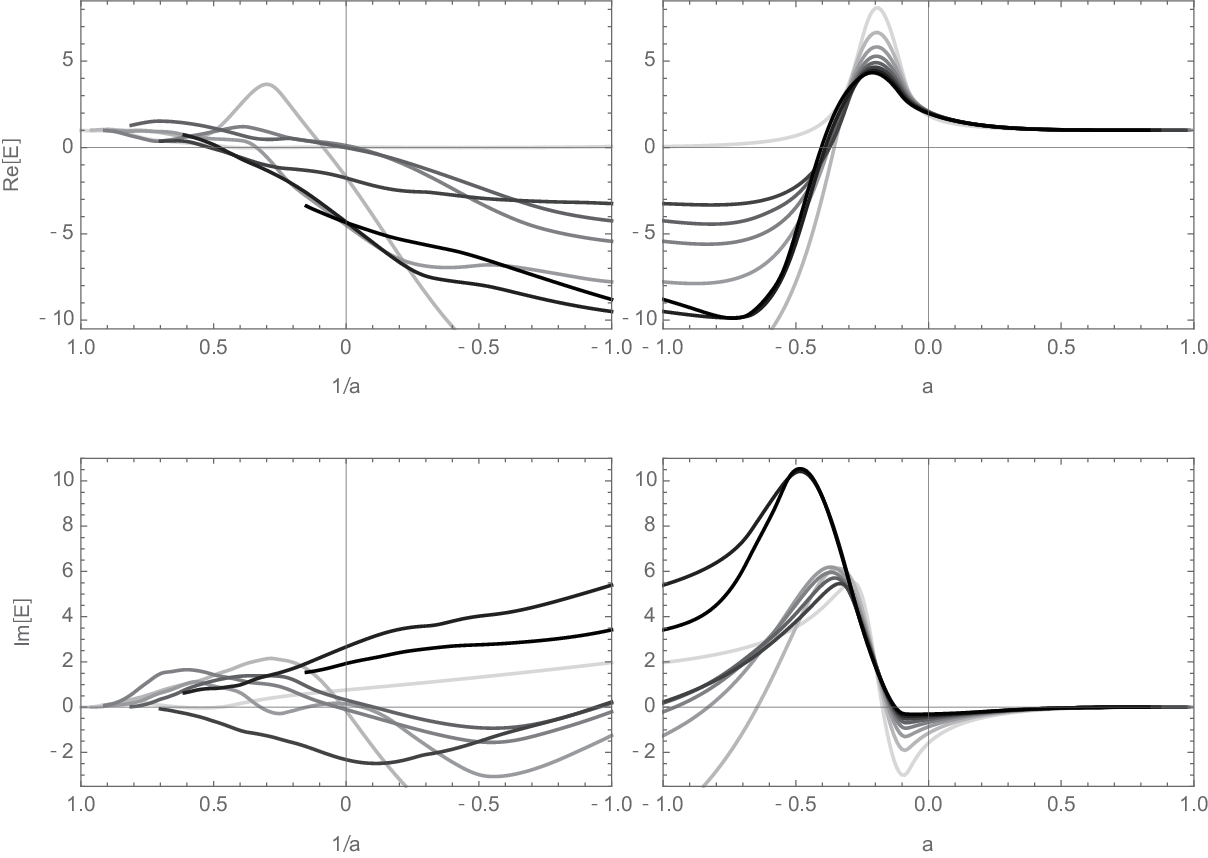}
\caption{Plots of the energy $E$ in Eq.~(\ref{eq:E-E0def}) of the ``double brane'' solution $\varPsi_{{\hat a}=1/a}^{\rm D}$ (left) and $\Psi_{a}^{\rm D}$ (right), truncated up to level $L=6,8,\dots,20$.
The upper (lower) figures show the real (imaginary) parts of $E$.
A darker line corresponds to a higher level. See also Table \ref{tab:a-range}.
\label{fig:ED}
}
\end{center}
\end{figure}

\begin{figure}[h]
\begin{center}
\includegraphics[width=16cm]{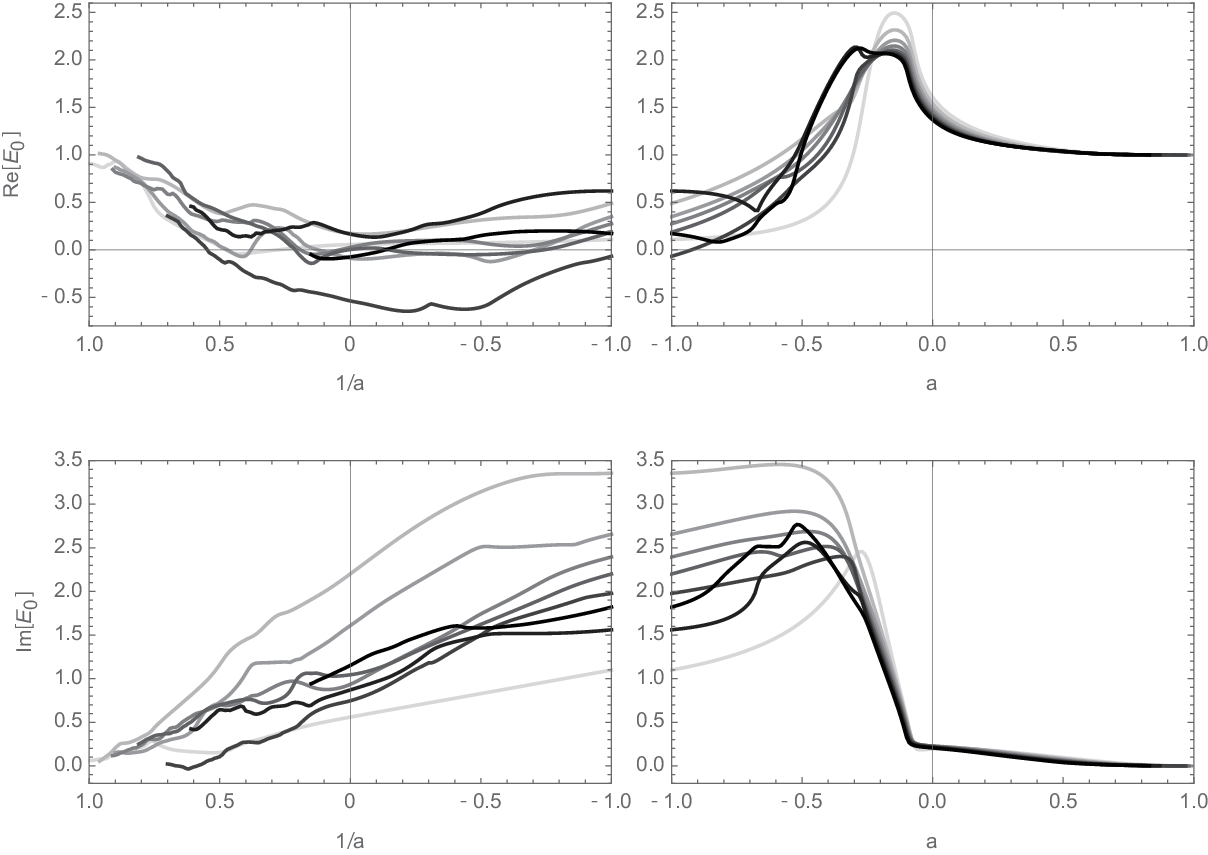}
\caption{Plots of the gauge-invariant observable $E_{0}$ from Eq.~(\ref{eq:E-E0def}) of the ``double brane'' solution $\varPsi_{{\hat a}=1/a}^{\rm D}$ (left) and $\Psi_{a}^{\rm D}$ (right), truncated up to level $L=6,8,\dots,20$.
The upper (lower) figures show the real (imaginary) parts of $E_{0}$.
A darker line corresponds to a higher level. See also Table \ref{tab:a-range}.
\label{fig:E0D}
}
\end{center}
\end{figure}

The ``double brane'' solution $\Psi_{a}^{\rm D}$ (or $\varPsi_{\hat a}^{\rm D}$) in various $a$-gauges has been constructed from the conventional $\Psi_{a=0}^{\rm D}$ in Siegel gauge as in Sect.~\ref{sec:initialconfig}. At the truncation level $L=20$, which is the highest one in our computation, $\Psi_{a}^{\rm D}$ was obtained for $-1\le a\le 0.83$, and $\varPsi_{\hat a}^{\rm D}$, which was constructed through the solution for $\hat{a}=1/a=-1$, was obtained for $-1\le \hat{a}\le 0.15$ (Table~\ref{tab:a-range}).
Plots of gauge invariants $E$ and $E_{0}$ from Eq.~(\ref{eq:E-E0def}) of the solutions $\varPsi^{\rm D}_{\hat a}$ and $\Psi^{\rm D}_{a}$ are given in Figs.~\ref{fig:ED} and \ref{fig:E0D}, where the real and imaginary parts of $E$ and $E_{0}$ are plotted separately.
We should note that $E$ and $E_{0}$ are complex in general because $\varPsi^{\rm D}_{\hat a}$ and $\Psi^{\rm D}_{a}$ do not satisfy the reality condition of the open string field in the same way as the ``double brane'' solution in Siegel gauge, at least for the truncation level $L\le 20$.

 The imaginary part of $E$ is nearly $0$ for $-0.1<a<1$.
For $a\sim 0$ (near Siegel gauge), the real part of $E$ is approximately $2$ 
as its name, ``double brane,'' suggests. However, $E\simeq 1$ for $0.5<a<1$ as in Fig.~\ref{fig:EDa0p5to1} in Appendix \ref{sec:detailednumericaldata}, which implies that the solution represents a single brane.
 As for $1>{\hat a}=1/a\ge -1$, $E$ varies significantly.
 Comparing Fig.~\ref{fig:E0D} with Fig.~\ref{fig:ED}, the gauge-invariant observable $E_{0}$ behaves similarly to the energy $E$ qualitatively.
 This is consistent with the expectation that $E=E_{0}$ for the solutions $\Psi_{a}^{\rm D}$ and $\varPsi_{\hat a}^{\rm D}$, although this does not hold quantitatively except for $0.5<a<1$, where $E_{0}$ approaches $1$ as in Fig.~\ref{fig:E0Da0p5to1} with the increasing truncation level in the same way as $E$ (Fig.~\ref{fig:EDa0p5to1}).
 
 The lowest level of the out-of-$a$-gauge equations $\Delta_{a}[\Psi^{\rm D}_{a}]$ approaches $0$ for $-0.1<a<1$ with the increasing truncation level, as in Fig.~\ref{fig:DeltaaD} in Appendix \ref{sec:validityagauge}. This is consistent with the validity of $\Psi_{a}^{\rm D}$ as a solution to the equation of motion in Eq.~(\ref{eq:EOM}). However, $\Delta_{a}[\Psi^{\rm D}_{a}]$ for $-1\le a<-0.5$ and $\Delta_{\hat a}[\varPsi^{\rm D}_{\hat a}]$ for $1\ge {\hat a}=1/a\ge -1$ do not seem to approach $0$ for higher level (Fig.~\ref{fig:DeltaaD}), which implies that the numerical solutions in these regions of $a$ might be inconsistent.
 
 From the ratio of the imaginary to real parts of the solution demonstrated in Fig.~\ref{fig:IoRD} in Appendix \ref{sec:validityagauge}, $\Psi^{\rm D}_{a}$ for $a\sim 0$ (near Siegel gauge) seems to approach a real solution with increasing truncation level, though $\Psi^{\rm D}_{a}$ tends to remain a complex one for $0.5<a<1$.

 These results might imply that, in the large-$L$ limit, $\Psi^{\rm D}_{a}$ for $a\sim 0$ becomes a real solution for double brane, and $\Psi^{\rm D}_{a}$ for $0.5<a<1$ becomes a complex solution for the single brane. By varying the value of $a$ from $0$ to $0.5$, $\Psi_{a}^{\rm D}$ may move onto a physically different branch of solutions.

\subsection{``Ghost brane'' solution
\label{sec:Gsol}
}

The ``ghost brane'' solution $\Psi_{a}^{\rm G}$ (or $\varPsi_{\hat a}^{\rm G}$) in various $a$-gauges has been constructed from the conventional $\Psi_{a=0}^{\rm G}$ in Siegel gauge as in Sect.~\ref{sec:initialconfig}.
At the truncation level $L=8$, $\Psi_{a}^{\rm G}$ was obtained for $-0.27\le a\le 1$ and $\varPsi_{\hat a}^{\rm G}$, which was constructed through the solution $\hat{a}=1/a=1$, was obtained for $1\ge \hat{a}\ge 0.95$. However, at the truncation levels $L=10,12,\dots,18$, the construction of 
$\Psi_{a}^{\rm G}$ reached neither $a=1$ nor $a=-1$.
At the truncation level $L=20$, which is the highest in our computation, $\Psi_{a}^{\rm G}$ was obtained for $-1\le a\le 0.83$.
Moreover, 
$\varPsi_{\hat a}^{\rm G}$, which was constructed through the solution for $\hat{a}=1/a=-1$, was obtained for $-1\le \hat{a}\le 0.28$ (Table~\ref{tab:a-range}).

\begin{figure}
\begin{center}
\includegraphics[width=16cm]{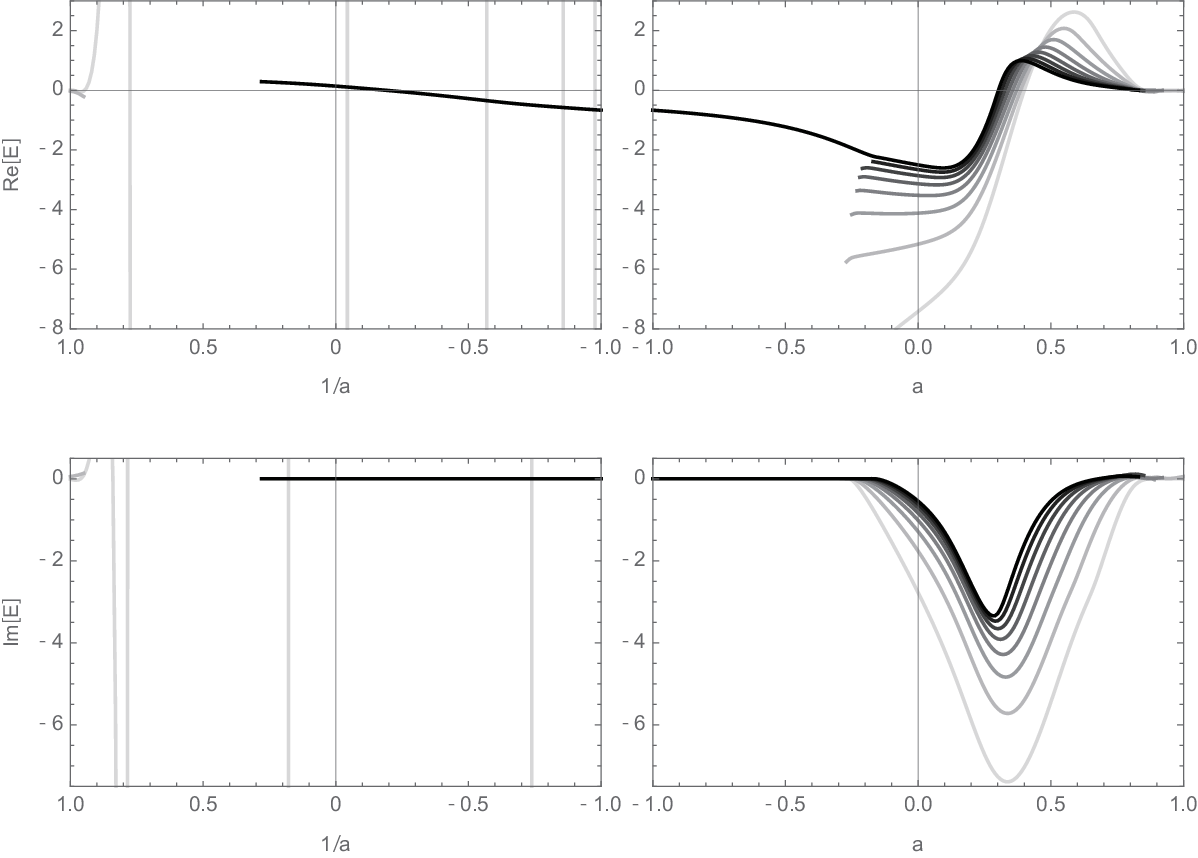}
\caption{Plots of the energy $E$ from Eq.~(\ref{eq:E-E0def}) of the ``ghost brane'' solution $\varPsi_{{\hat a}=1/a}^{\rm G}$ (left) and $\Psi_{a}^{\rm G}$ (right), truncated up to level $L=6,8,\dots,20$.
The upper (lower) figures show the real (imaginary) parts of $E$.
A darker line corresponds to a higher level.
\label{fig:EG}
}
\end{center}
\end{figure}

\begin{figure}
\begin{center}
\includegraphics[width=16cm]{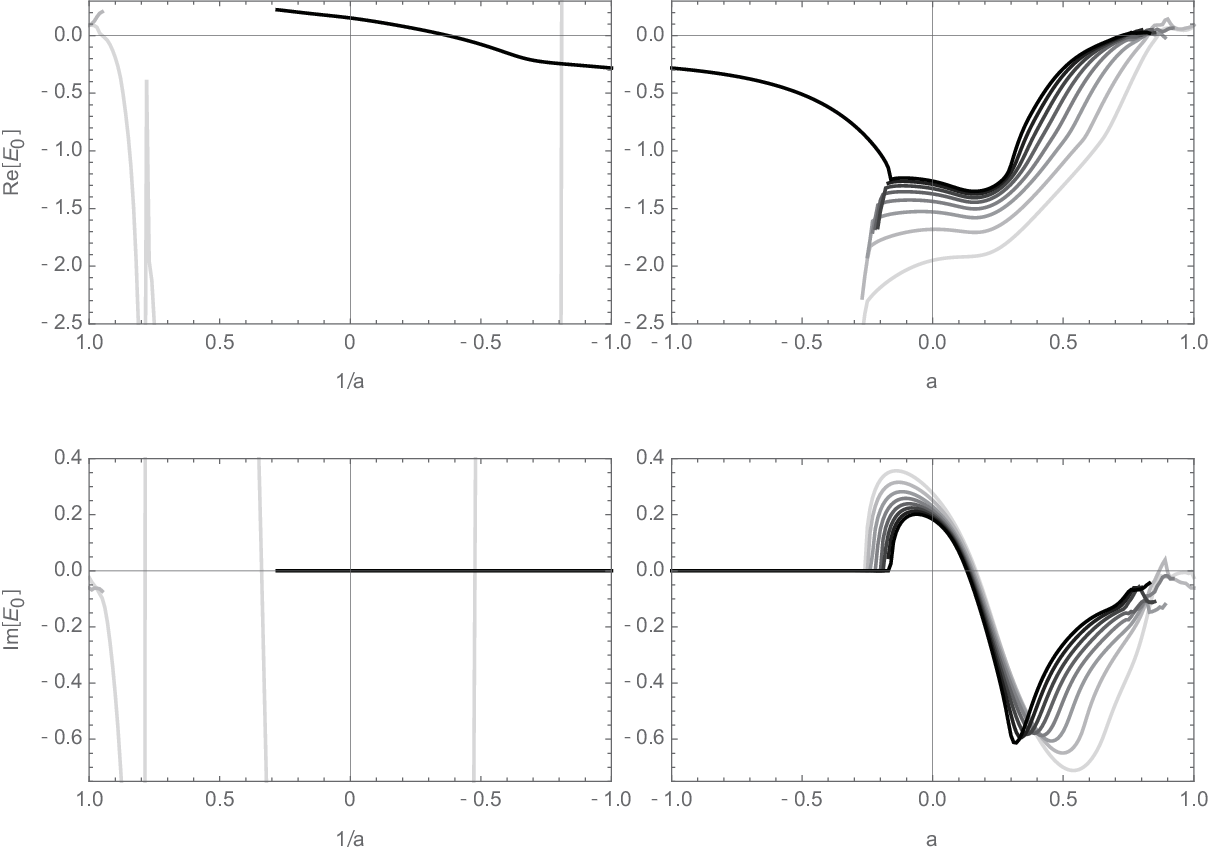}
\caption{Plots of the gauge-invariant observable $E_{0}$ in Eq.~(\ref{eq:E-E0def}) of the ``ghost brane'' solution $\varPsi_{{\hat a}=1/a}^{\rm G}$ (left) and $\Psi_{a}^{\rm G}$ (right), truncated up to level $L=6,8,\dots,20$.
The upper (lower) figures show the real (imaginary) parts of $E_{0}$.
A darker line corresponds to a higher level.
\label{fig:E0G}
}
\end{center}
\end{figure}

Plots of the gauge invariants $E$ and $E_{0}$ of the solutions are shown in Figs.~\ref{fig:EG} and \ref{fig:E0G}, where the real and imaginary parts of $E$ and $E_{0}$ are plotted separately.
We note that $E$ and $E_{0}$ are complex in general, because $\varPsi^{\rm G}_{\hat a}$ and $\Psi^{\rm G}_{a}$ do not satisfy the reality condition of the open string field in the same way as the ``double brane'' solution for truncation levels $L\le 18$. At the truncation level $L=20$, $\Psi_{a}^{\rm G}$ for $-1\le a\le -0.17$  and $\varPsi_{\hat a}^{\rm G}$ for $-1\le {\hat a}\le 0.28$ satisfy the reality condition, as is shown in Fig.~\ref{fig:IoRG} in Appendix \ref{sec:validityagauge}. Namely, $\Psi_{a}^{\rm G}$ becomes real at $a=-0.17$ when the value of $a$ varies from $0$ to $-1$. 

The energy $E$ is approximately $-2.5$ for $a\sim 0$ (near Siegel gauge). The real part of $E$ varies from $-2.5$ to $0$ for $0<a<1$.
The imaginary part of $E$ seems to approach $0$ with increasing truncation level.
Comparing Fig.~\ref{fig:E0G} with Fig.~\ref{fig:EG},
the real part of $E_{0}$ behaves similarly to that of $E$ qualitatively, although they are rather different quantitatively.
The behavior of the imaginary part of $E_{0}$ is somewhat similar to that of $E$.


As in Fig.~\ref{fig:DeltaaG} in Appendix \ref{sec:validityagauge}, the absolute value of the lowest level of the out-of-$a$-gauge equations $|\Delta_{a}[\Psi_{a}^{\rm G}]|$ is decreasing with increasing truncation level.
This shows consistency of $\Psi_{a}^{\rm G}$ with the equation of motion in Eq.~(\ref{eq:EOM}).
Although the ratio of the imaginary part to the real one of $\Psi_{a}^{\rm G}$ is decreasing for $a\sim 0.5$ with increasing truncation level as in Fig.~\ref{fig:IoRG}, it may not vanish for $0\le a<1$ in the large-$L$ limit.
We note that, according to Ref.~\cite{Kudrna:2018mxa}, $\Psi_{a=0}^{\rm G}$ in Siegel gauge does not satisfy the reality condition from the extrapolation to $L=\infty$.
Because $\Psi_{a}^{\rm G}$ for $-1\le a\le -0.17$ at the truncation level $L=20$ is real, as mentioned, $\Psi_{a}^{\rm G}$ might move onto a branch of real solutions from that of complex ones when the value of $a$ varies from $0$ to $-1$ at $L=\infty$.
 
\section{Concluding remarks
 \label{sec:remarks}}
 
 We have constructed numerical solutions in $a$-gauge starting from the solutions in Siegel gauge for tachyon vacuum, ``double brane,'' and ``ghost brane.''
 We solved the equation of motion obtained from the $a$-gauge fixed action in the twist-even universal space, and carried out calculations up to truncation level $L=20$.
 
The solution for the tachyon vacuum in $a$-gauge, $\Psi_{a}^{\rm T}$ or $\varPsi_{{\hat a}=1/a}^{\rm T}$, has energy $E\simeq 0$ and the gauge-invariant observable $E_{0}\simeq 0$ for almost all $a$, as we expected from the known results.
This is consistent with the physical interpretation as the solution for no brane.

The solution for the ``double brane'' in $a$-gauge, $\Psi_{a}^{\rm D}$, has $E\sim 2$ for $a\sim 0$ and $E\simeq 1$ for $0.5<a<1$,
which we did not expect because $E$ is gauge invariant.
At least naively, we can interpret this as $\Psi_{a}^{\rm D}$ representing a double brane near Siegel gauge and a single brane for $0.5<a<1$. 
Namely, we may be able to expect that, for a particular value $a_{\rm c}$, $\Psi_{a}^{\rm D}$ corresponds to a real solution for the double brane in the range $0\le a<a_{\rm c}$ and a complex solution for a single brane in the range $a_{\rm c}<a<1$ in the large-$L$ limit. At $a=a_{\rm c}$, $\Psi_{a}^{\rm D}$ might jump to a different branch of solutions.
Alternatively, there might be a possibility that $\Psi_{a}^{\rm D}$ corresponds to a single brane for $0\le a<1$, taking into account the behavior of the gauge-invariant observable $E_{0}$, because $E_{0}\sim 1.4$ for $a\sim 1$.
Originally, such a possibility was mentioned in Siegel gauge in Ref.~\cite{Kudrna:2018mxa}.
It is necessary to investigate the theory around the solution $\Psi_{a}^{\rm D}$ to understand its physical meaning.\footnote{
The spectrum was investigated for the theory around the tachyon vacuum solution in Siegel gauge in Refs.~\cite{Giusto:2003wc,Imbimbo:2006tz}. 
Such analysis may be useful for other solutions.
Or, it is desirable to gain an insight from analytic multi-brane solutions; see Refs.~\cite{Murata:2011ep,Hata:2015aoh, Erler:2014eqa} and subsequent works.
}
As for the other region of $a$, $\Psi_{a}^{\rm D}$ for $-1\le a<-0.5$ and $\varPsi_{\hat a}^{\rm D}$ for $1>{\hat a}=1/a\ge -1$ seem to be inconsistent with the equation of motion in Eq.~(\ref{eq:EOM}).

The solution for the ``ghost brane'' in $a$-gauge, $\Psi_{a}^{\rm G}$, has $E\sim -2.5$ for $a\sim 0$ and $E\sim 0$ for $a\sim 0.8$,
which we did not expect in the same manner as the ``double brane'' solution.
Although $\Psi_{a}^{\rm G}$ seems to be complex for $0\le a<1$ in the large-$L$ limit, it becomes real for $a\le -0.17$ at $L=20$.
Physical interpretation of the ``ghost brane'' solution is more complicated than the ``double brane'' solution.
 
\begin{figure}[htbp]
\begin{center}
\includegraphics[width=16cm]{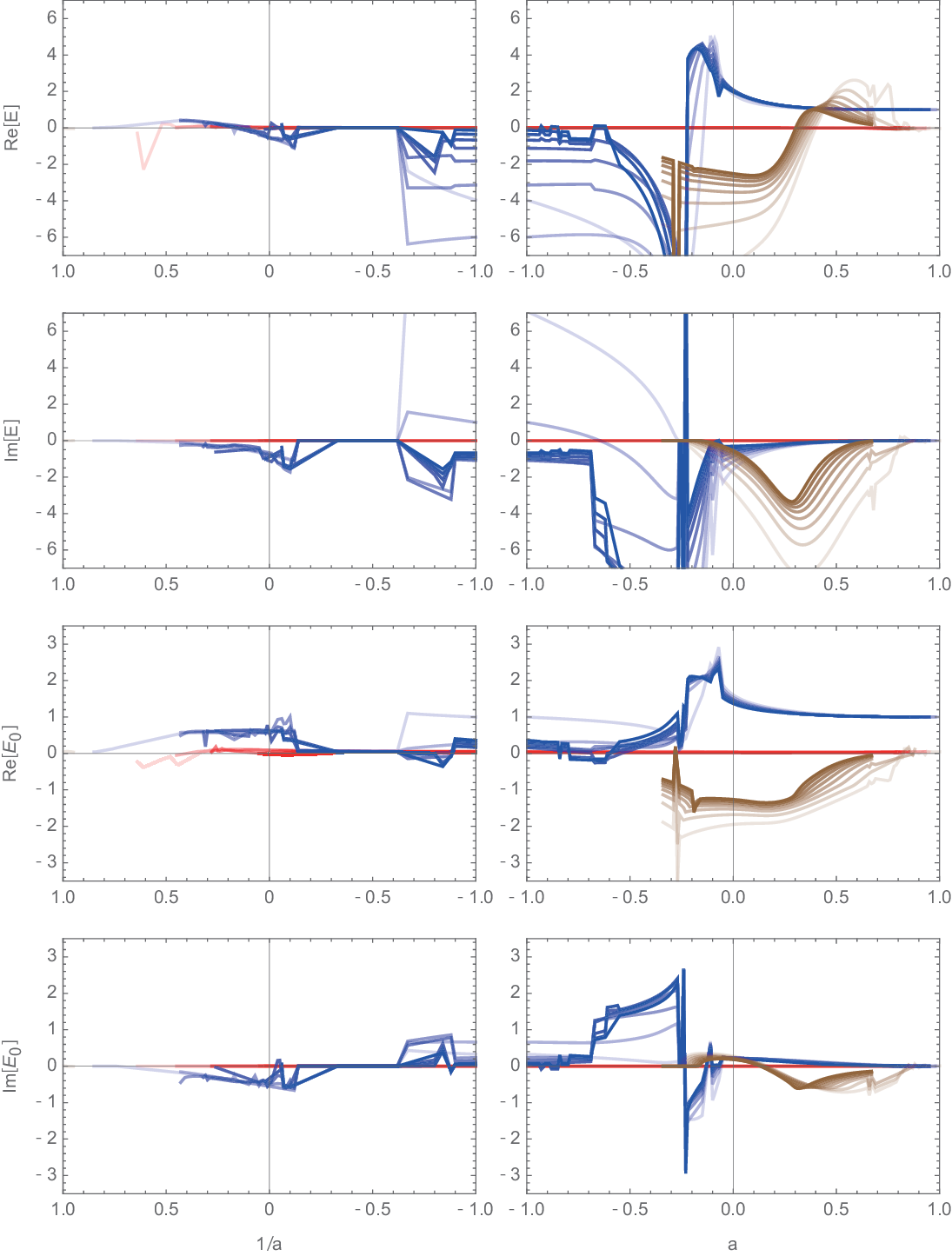}\\
\caption{Plots of $E$ and $E_{0}$ from Eq.~(\ref{eq:E-E0def}) of 
$\varPsi_{{\hat a}=1/a}^{{\rm T}\prime},\Psi_{a}^{{\rm T}\prime}$ (red),
$\varPsi_{{\hat a}=1/a}^{{\rm D}\prime},\Psi_{a}^{{\rm D}\prime}$ (blue), and 
$\varPsi_{{\hat a}=1/a}^{{\rm G}\prime},\Psi_{a}^{{\rm G}\prime}$ (brown),
truncated up to level $L=6,8,\dots,20$.
A darker color corresponds to a higher level.
\label{fig:EE0TGDL}}
\end{center}
\end{figure}

 In order to solve the equation of motion in $a$-gauge numerically, we divided the region of $a$ in two: $-1\le a\le 1$ and $-1\le {\hat a}=1/a\le 1$.
We constructed solutions in $a$-gauge, $(\Psi_{a}^{\rm T},\Psi_{a}^{\rm D},\Psi_{a}^{\rm G})$ and $(\varPsi_{\hat a}^{\rm T},\varPsi_{\hat a}^{\rm D},\varPsi_{\hat a}^{\rm G})$, from known solutions in Siegel gauge, $(\Psi_{a=0}^{\rm T},\Psi_{a=0}^{\rm D},\Psi_{a=0}^{\rm G})$, by varying the value of $a$ little by little ($\varDelta a=\pm 0.01$ and $\varDelta \hat{a}=\mp 0.01$) at each truncation level $L$ (Sect.~\ref{sec:initialconfig}).
Alternatively, we also constructed solutions as follows: we first construct solutions in $a$-gauge from $a=0$ by varying the value of $a$ little by little ($\varDelta a=\pm 0.01$ and $\varDelta \hat{a}=\mp 0.01$) at the lowest truncation level, and then we construct solutions level by level at each fixed value of $a$.
Here, we denote such solutions with prime: $(\Psi_{a}^{{\rm T}\prime},\Psi_{a}^{{\rm D}\prime},\Psi_{a}^{{\rm G}\prime})$ and $(\varPsi_{\hat a}^{{\rm T}\prime},
\varPsi_{\hat a}^{{\rm D}\prime},\varPsi_{\hat a}^{{\rm G}\prime})$.
Plots of the energy and the gauge-invariant observable for them are shown in Fig.~\ref{fig:EE0TGDL}. There are many spiky points in Fig.~\ref{fig:EE0TGDL} because we could not obtain solutions from lower levels for some values of $a$.
We note that we joined adjacent data points, where solutions exist, with line segments at each truncation level.
Comparing Fig.~\ref{fig:EE0TGDL} with Figs.~\ref{fig:ETGD} and \ref{fig:E0TGD}, we can easily find that the values of $E$ and $E_{0}$ of $(\Psi_{a}^{{\rm D}\prime},\Psi_{a}^{{\rm G}\prime})$ and $(\varPsi_{\hat a}^{{\rm D}\prime},\varPsi_{\hat a}^{{\rm G}\prime})$
are different from those of $(\Psi_{a}^{{\rm D}},\Psi_{a}^{{\rm G}})$ and $(\varPsi_{\hat a}^{{\rm D}},\varPsi_{\hat a}^{{\rm G}})$ in some regions of $a$, respectively.
The values of $E$ and $E_{0}$ of $\varPsi_{\hat a}^{{\rm T}\prime}$ are also different from those of $\varPsi_{\hat a}^{{\rm T}}$ at some points of $a$, although they are approximately $0$ for all $a$ at higher truncation levels. (For example, see Tables~\ref{tab:EsampleT} and \ref{tab:E0sampleT} in Appendix \ref{sec:detailednumericaldata}.)
In general, there is a possibility that $(\Psi_{a}^{\rm T},\Psi_{a}^{\rm D},\Psi_{a}^{\rm G})$ and $(\Psi_{a}^{{\rm T}\prime},\Psi_{a}^{{\rm D}\prime},\Psi_{a}^{{\rm G}\prime})$ are in different branches of solutions at the same value of $a$ except for $a=0$ (Siegel gauge). Furthermore, it is more possible that $(\varPsi_{\hat a}^{\rm T},\varPsi_{\hat a}^{\rm D},\varPsi_{\hat a}^{\rm G})$ and $(\varPsi_{\hat a}^{{\rm T}\prime},\varPsi_{\hat a}^{{\rm D}\prime},\varPsi_{\hat a}^{{\rm G}\prime})$ are in different branches at the same value of ${\hat a}$.
Therefore, it seems to be difficult to find a plausible method for extrapolations to $L=\infty$ in $a$-gauge for $a\ne 0$.

So far, we have investigated numerical solutions with reference to those in Siegel gauge.
It is an interesting problem to find other physically meaningful solutions in ($a=\infty$)-gauge or, more generally, $(a\ne 0)$-gauge.

\section*{Acknowledgments}
We would like to thank T.~Takahashi for valuable comments.
The author thanks the Yukawa Institute for Theoretical Physics at Kyoto University.
Discussions during the YITP workshop YITP-W-21-04 on ``Strings and Fields 2021'' were useful in completing this work.
This work was supported in part by JSPS KAKENHI Grant Numbers
JP20K03933, JP20K03972. The numerical calculations were partly carried
out on sushiki and Yukawa-21 at YITP in Kyoto University.

\appendix
\section{Some numerical data for solutions
\label{sec:App}
}

\subsection{State space dimension
\label{sec:dimension}
}

In the $(b,c)$-ghost sector for the ghost number $g$ at the level
$\ell$, we can take a basis of the form
\begin{align}
&b_{-n_1}\cdots b_{-n_k}c_{-m_1}\cdots c_{-m_{k^{\prime}}}c_1|0\rangle_{\rm g},
\qquad n_1 > \cdots > n_k\ge 1,\qquad
m_1 > \cdots > m_{k^{\prime}}\ge 0,\nonumber\\
&k^{\prime}-k+1=g,\qquad
\sum_{i=1}^kn_i+\sum_{j=1}^{k^{\prime}}m_j=\ell.
\label{eq:bc-basis}
\end{align}
A generating function of its dimension ${\tilde d}_{g,\ell}$ is given by \cite{Kudrna:2018mxa, Rastelli:2000iu}
\begin{align}
\sum_{g\in \mathbb{Z}}\sum_{\ell=0}^{\infty}\tilde d_{g,\ell}\, q^{\ell}y^g=y(1+y)
\prod_{n=1}^{\infty}\left\{(1+q^ny)(1+q^ny^{-1})\right\}
=\prod_{n=1}^{\infty}(1-q^n)^{-1}\,\sum_{g\in\mathbb{Z}}y^g q^{\frac{1}{2}(g-1)(g-2)}.
\end{align}
In Siegel gauge, ``$m_1 > \cdots > m_{k^{\prime}}\ge 0$'' in Eq.~(\ref{eq:bc-basis}) is replaced by ``$m_1 > \cdots > m_{k^{\prime}}\ge 1$'' and a generating function of its dimension $d_{g,\ell}$ is
\begin{align}
\sum_{g\in \mathbb{Z}}\sum_{\ell=0}^{\infty}d_{g,\ell}\, q^{\ell}y^g&=
y\prod_{n=1}^{\infty}\left\{(1+q^ny)(1+q^ny^{-1})\right\}\nonumber\\
&=\prod_{n=1}^{\infty}(1-q^n)^{-1}\,\sum_{g\in\mathbb{Z}}y^g
\sum_{k=0}^{\infty}\left(
q^{\frac{1}{2}(|g-1|+2k)(|g-1|+2k+1)}-q^{\frac{1}{2}(|g-1|+2k+1)(|g-1|+2k+2)}
\right).
\label{eq:dgelldef}
\end{align}
Noting that
\begin{align}
\sum_{\ell=0}^{\infty}(d_{g,\ell}+d_{3-g,\ell})q^{\ell}
&=\prod_{n=1}^{\infty}(1-q^n)^{-1}
\sum_{k=0}^{\infty}\Bigl(
q^{\frac{1}{2}(|g-1|+2k)(|g-1|+2k+1)}-q^{\frac{1}{2}(|g-1|+2k+1)(|g-1|+2k+2)}
\nonumber\\
&\qquad\qquad\qquad\qquad
+q^{\frac{1}{2}(|2-g|+2k)(|2-g|+2k+1)}-q^{\frac{1}{2}(|2-g|+2k+1)(|2-g|+2k+2)}
\Bigr)
\nonumber\\
&=q^{\frac{1}{2}(g-1)(g-2)}\prod_{n=1}^{\infty}(1-q^n)^{-1}
=\sum_{\ell=0}^{\infty}\tilde{d}_{g,\ell}q^{\ell}=\sum_{\ell=0}^{\infty}\tilde{d}_{3-g,\ell}q^{\ell},
\end{align}
we have relations
\begin{align}
\tilde d_{g,\ell}=\tilde d_{3-g,\ell}=d_{g,\ell}+d_{3-g,\ell}.
\label{eq:ddrelation}
\end{align}
In the matter sector of the universal state space at level $\ell$, 
we have a basis of the form
\begin{align}
&L_{-l_1}^{\rm mat}\cdots L_{-l_m}^{\rm mat}|0\rangle_{\rm m},
&&l_1\ge \cdots\ge l_m\ge 2,\qquad\qquad
\sum_{i=1}^ml_i=\ell.
\end{align}
A generating function of its dimension $d_{\ell}$ is
\begin{align}
\sum_{\ell=0}^{\infty}d_{\ell}q^{\ell}=\prod_{n=2}^{\infty}(1-q^n)^{-1}.
\label{eq:delldef}
\end{align}
We denote the dimension of the universal state space in Siegel gauge for the ghost number $g$ at level $\ell$ as $D_{g,\ell}$ and it is given by
\begin{align}
D_{g,\ell}=\sum_{k=0}^{\ell}d_{\ell-k}d_{g,k}.
\label{eq:Dgelldef}
\end{align}
In particular, the dimension of the universal state space in Siegel gauge for the ghost number $1$ at level $\ell$, which does not contain $c_{0}$, is $D_{1,\ell}$, and the dimension of remaining part of the universal state space for the ghost number $1$ given by
 \begin{align}
 \sum_{k=0}^{\ell}d_{\ell-k}({\tilde d}_{1,k}-d_{1,k})=\sum_{k=0}^{\ell}d_{\ell-k}d_{2,k}=D_{2,\ell},
 \end{align}
 where use has been made of Eq.~(\ref{eq:ddrelation}).
 We list $d_{\ell}$, $d_{1,\ell}$, $d_{2,\ell}$, $D_{1,\ell}$, and $D_{2,\ell}$ in Table~\ref{tab:d1Ld2L} up to level $20$.
 The dimension of the twist-even universal state space in Siegel gauge truncated up to level $L$ for the ghost number $g$ is
 \begin{align}
N_{g,L}&=\sum_{l=0}^{L/2}D_{g,2l}.
\label{eq:NgLdef}
 \end{align}
The number of arbitrary coefficients of a string field with ghost number $1$ in $a$-gauge, $t_{i_{l}}$ in $\Psi_{a}$, Eq.~(\ref{eq:Psi_adef}), or $(v_{i^{\prime}_{l}},u_{b_{l}})$ in $\varPsi_{\hat a}$, Eq.~(\ref{eq:varPsiahat}), is given by $N_{1,L}$.
The number of the out-of-$a$-gauge equations, Eqs.~(\ref{eq:outofgauge}) or (\ref{eq:outofgaugehat}), truncated up to level $L$ is $N_{2,L}$.
They are listed in Table~\ref{tab:univedim}.
\begin{table}[htbp]
  \begin{minipage}[t]{.61\textwidth}
         \caption{
         The dimensions of the matter universal state space at level $\ell$, $d_{\ell}$ from Eq.~(\ref{eq:delldef}); 
         the ghost universal state space in Siegel gauge at level
         $\ell$, 
         $d_{g,\ell}$ from Eq.~(\ref{eq:dgelldef}), for the ghost number $g=1,2$; 
         and the universal state space in Siegel gauge at level $\ell$, 
         $D_{g,\ell}$ from Eq.~(\ref{eq:Dgelldef}), for the ghost number $g=1,2$.
         \label{tab:d1Ld2L}}
    \begin{center}
   \renewcommand{\arraystretch}{0.85}
         \begin{tabular}{r|r|r|r|r|r}
\hline
$\ell$&$d_{\ell}$&$d_{1,\ell}$&$d_{2,\ell}$&$D_{1,\ell}$& $D_{2,\ell}$
\\
\hline\hline
 0&1&1&0&1&0\\
1&0&0&1&0&1\\
2&1&1&1&2&1\\
3&1&2&1&3&2\\
4&2&3&2&6&4\\
5&2&4&3&9&7\\
6&4&6&5&17&12\\
7&4&8&7&25&20\\
8&7&12&10&43&32\\
9&8&16&14&64&51\\
10&12&23&19&102&79\\
11&14&30&26&150&121\\
12&21&42&35&231&182\\
13&24&54&47&333&272\\
14&34&73&62&496&399\\
15&41&94&82&709&582\\
16&55&124&107&1027&839\\
17&66&158&139&1448&1200\\
18&88&206&179&2060&1700\\
19&105&260&230&2866&2394\\
20&137&334&293&4010&3342\\
\hline
      \end{tabular}
\renewcommand{\arraystretch}{1}
    \end{center}
    \end{minipage}
\hfill
\begin{minipage}[t]{.34\textwidth}
        \caption{
        Dimension of the twist-even universal state space in Siegel gauge truncated up to level $L$, $N_{g,L}$ from Eq.~(\ref{eq:NgLdef}), for the ghost number $g=1,2$.
            \label{tab:univedim}}
    \begin{center}
     \renewcommand{\arraystretch}{1.15}
      \begin{tabular}{r|r|r}
\hline
$L$&$N_{1,L}$&$N_{2,L}$\\
\hline\hline
0&1&0\\
2&3&1\\
4&9&5\\
6&26&17\\
8&69&49\\
10&171&128\\
12&402&310\\
14&898&709\\
16&1925&1548\\
18&3985&3248\\
20&7995&6590\\
\hline
      \end{tabular}
       \renewcommand{\arraystretch}{1}
    \end{center}
   \end{minipage}
\end{table}

\subsection{Some detailed numerical data for solutions
\label{sec:detailednumericaldata}
}

We tried to solve Eq.~(\ref{eq:EOM_AK}) for $-1\le a\le 1$, and Eqs.~(\ref{eq:EOMhata1}) and (\ref{eq:EOMhata2}) for $-1\le \hat{a}=1/a\le 1$, using Newton's method with the initial configurations chosen as in Sect.~\ref{sec:initialconfig}.
We were able to obtain solutions in $a$-gauge, $\Psi_{a}^{\rm T}$ and $\varPsi_{\hat a}^{\rm T}$ for tachyon vacuum, $\Psi_{a}^{\rm D}$ and $\varPsi_{\hat a}^{\rm D}$ for ``double brane,'' and $\Psi_{a}^{\rm G}$ and $\varPsi_{\hat a}^{\rm G}$ for ``ghost brane,''  in respective ranges in $-1\le a\le 1$ and $-1\le \hat{a}\le 1$ at each truncation level $L$.
We list them in Table~\ref{tab:a-range}.\\

\begin{table}[htbp]
\caption{The ranges of $\hat{a}=1/a$ and $a$ where numerical solutions $\varPsi_{\hat{a}}$ or $\Psi_{a}$ in Sect.~\ref{sec:initialconfig} were obtained.
$L$ is the truncation level.
Except for $\varPsi^{\rm G}_{\hat a}$ at $L=4,6,8$,
$\varPsi_{\hat{a}}$ were constructed through $\hat{a}=1/a=-1$.
\label{tab:a-range}
}
\begin{center}
\renewcommand{\arraystretch}{1.1}
\scalebox{0.9}{
\begin{tabular}{r||c|c||c|c||c|c}
\hline
$L$&$\varPsi^{\rm T}_{\hat a}$&$\Psi^{\rm T}_{a}$&
$\varPsi^{\rm D}_{\hat a}$&$\Psi^{\rm D}_{a}$&
$\varPsi^{\rm G}_{\hat a}$&$\Psi^{\rm G}_{a}$\\
\hline
$2$&
$0.65\ge \hat{a}\ge -1$&$-1\le a\le 1$&
$1\ge \hat{a}\ge -1$&$-1\le a \le 0.99$&
 --- & ---
\\
\hline
$4$&
$0.52\ge \hat{a}\ge -1$&$-1\le a\le 0.97$&
$1\ge \hat{a}\ge -1$&$-1\le a\le 0.99$&
$1\ge \hat{a}\ge 0.9$&$-0.34\le a\le 1$
\\
\hline
$6$&
$0.35\ge \hat{a}\ge -1$&$-1\le a\le 0.86$&
$1\ge \hat{a}\ge -1$&$-1\le a\le 0.99$&
$1\ge \hat{a}\ge -1$&$-0.29\le a\le 1$
\\
\hline
$8$&
$0.46\ge \hat{a}\ge -1$&$-1\le a\le 0.91$&
$0.96\ge\hat{a}\ge -1$&$-1\le a\le 0.99$&
$1\ge\hat{a}\ge 0.95$&$-0.27\le a\le 1$
\\
\hline
$10$&
$0.24\ge \hat{a}\ge -1$&$-1\le a\le 0.82$&
$0.9\ge \hat{a}\ge -1$&$-1\le a\le 0.98$&
---&$-0.25\le a\le 0.89$
\\
\hline
$12$&
$0.41\ge\hat{a}\ge -1$&$-1\le a\le 0.82$&
$0.91\ge \hat{a}\ge -1$&$-1\le a\le 0.98$&
---&$-0.23\le a\le 0.92$
\\
\hline
$14$&
$0.2\ge\hat{a}\ge -1$&$-1\le a\le 0.81$&
$0.81\ge\hat{a}\ge -1$&$-1\le a\le 0.96$&
---&$-0.22\le a\le 0.84$
\\
\hline
$16$&
$0.44\ge\hat{a}\ge -1$&$-1\le a\le 0.81$&
$0.7\ge\hat{a}\ge -1$&$-1\le a\le 0.97$&
---&$-0.21\le a\le 0.85$
\\
\hline
$18$&
$0.11\ge\hat{a}\ge -1$&$-1\le a\le 0.79$&
$0.61\ge \hat{a}\ge -1$&$-1\le a\le 0.87$&
---&$-0.17\le a\le 0.8$
\\
\hline
$20$&
$0.23\ge \hat{a}\ge -1$&$-1\le a\le 0.79$&
$0.15\ge\hat{a}\ge -1$&$-1\le a\le 0.83$&
$0.28\ge\hat{a}\ge -1$&$-1\le a\le 0.83$
\\
\hline
\end{tabular}
}
\renewcommand{\arraystretch}{1}
\end{center}
\end{table}

We show explicit values of the energy $E$ and the gauge-invariant observable $E_{0}$ for the tachyon vacuum solution in $a$-gauge for some values of $a$ in Tables~\ref{tab:EsampleT} and \ref{tab:E0sampleT}.
They correspond to updates of the results in Refs.~\cite{Asano:2006hm,Kishimoto:2009cz},
where $\pm (E-1)$ and $1-E_{0}$ were listed in tables.
We note that a different method from that this paper used here was adopted in Ref.~\cite{Kishimoto:2009cz} to solve the equation of motion in $a$-gauge,
which causes slight discrepancies in the resulting values.
In Tables~\ref{tab:EsampleT} and \ref{tab:E0sampleT}, it may look curious that $E$ and $E_{0}$ for $L=10,14$ in ($a=4$)-gauge are not available, although those for $L=12,16$ exist.
This is because we have constructed solutions from those in Siegel gauge ($a=0$) at each truncation level $L$ as in Sect.~\ref{sec:initialconfig}.
Moreover, we constructed solutions for the tachyon vacuum level by level in ($a=4$)-gauge and obtained them up to truncation level $L=12$.
We found that such solutions, which we denote as $\varPsi_{\hat{a}=1/4}^{{\rm T}\prime}$, give $E[\varPsi_{\hat{a}=1/4}^{{\rm T}\prime}]=0.0710907\,(L=10)$ and $0.0742312\,(L=12)$, and $E_{0}[\varPsi_{\hat{a}=1/4}^{{\rm T}\prime}]=0.0939597\,(L=10)$ and $0.0673891\,(L=12)$,
where the values for $L=12$ are different from those in Tables~\ref{tab:EsampleT} and \ref{tab:E0sampleT}.
This implies that $\varPsi_{\hat{a}=1/4}^{{\rm T}}$ and $\varPsi_{\hat{a}=1/4}^{{\rm T}\prime}$ are in different branches of solutions to Eqs.~(\ref{eq:EOMhata1}) and  (\ref{eq:EOMhata2}) at $L=12$.
Except for them, the values in Tables~\ref{tab:EsampleT} and \ref{tab:E0sampleT} for the solutions constructed as in Sect.~\ref{sec:initialconfig} coincide with those of solutions $\Psi_{a}^{{\rm T}\prime}$ and $\varPsi_{\hat a}^{{\rm T}\prime}$ obtained level by level at each fixed value of $a$, although we could not obtain $\varPsi_{{\hat a}=0}^{{\rm T}\prime}$ for $L=18,20$.
\\

\begin{table}[htbp]
\caption{The energy $E$ from Eq.~(\ref{eq:E-E0def}) of the tachyon vacuum solution $\varPsi_{\hat a}^{\rm T}$ for $\hat{a}=1/a=1/4,0,-1/2$ and $\Psi_{a}^{\rm T}$ for $a=0,1/2$, truncated up to level $L$. ``--'' corresponds to a solution which was not obtained as in Table~\ref{tab:a-range}.
\label{tab:EsampleT}
}
\begin{center}
\renewcommand{\arraystretch}{1.1}
\begin{tabular}{r||l|l|l|l|l}
\hline
$L$&$a=4$&$a=\infty$&$a=-2$&$a=0$&$a=1/2$\\
\hline
$2$&$0.114805$&$0.0867229$&$0.0652269$&$0.0406234$&$0.022964$\\
\hline
$4$&$0.0881147$&$0.0524154$&$0.031577$&$0.0121782$&$-0.000301875$\\
\hline
$6$&$0.0755114$&$0.0390562$&$0.0201057$&$0.00482288$&$-0.00458577$\\
\hline
$8$&$0.0710871$&$0.0314585$&$0.0144505$&$0.00206982$&$-0.00548353$\\
\hline
$10$&\qquad --&$0.0261408$&$0.0110531$&$0.000817542$&$-0.00551823$\\
\hline
$12$&$0.0429279$&$0.0160149$&$0.00874361$&$0.000177737$&$-0.00530578$\\
\hline
$14$&\qquad --&$-0.002015$&$0.00703121$&$-0.00017373$&$-0.00503144$\\
\hline
$16$&$0.00584047$&$-0.00730175$&$0.00567063$&$-0.000375452$&$-0.00475668$\\
\hline
$18$&\qquad --&$-0.00924349$&$0.00451818$&$-0.000493711$&$-0.00450181$\\
\hline
$20$&\qquad --&$-0.00816331$&$0.00347154$&$-0.000562955$&$-0.00427228$\\
\hline
\end{tabular}
\renewcommand{\arraystretch}{1}
\end{center}
\end{table}
\begin{table}[htbp]
\caption{The gauge invariant-observable $E_{0}$ from Eq.~(\ref{eq:E-E0def}) of the tachyon vacuum solution $\varPsi_{\hat a}^{\rm T}$ for $\hat{a}=1/a=1/4,0,-1/2$ and $\Psi_{a}^{\rm T}$ for $a=0,1/2$, truncated up to level $L$. ``--''  corresponds to a solution which was not obtained as in Table~\ref{tab:a-range}.
\label{tab:E0sampleT}
}
\begin{center}
\renewcommand{\arraystretch}{1.1}
\begin{tabular}{r||l|l|l|l|l}
\hline
$L$&$a=4$&$a=\infty$&$a=-2$&$a=0$&$a=1/2$\\
\hline
$2$&$0.101143$&$0.12419$&$0.120745$&$0.110138$&$0.100313$\\
\hline
$4$&$0.114702$&$0.114305$&$0.0912254$&$0.0680476$&$0.0568528$\\
\hline
$6$&$0.094224$&$0.0942222$&$0.0684689$&$0.0489211$&$0.0424566$\\
\hline
$8$&$0.100325$&$0.0939022$&$0.0599348$&$0.0388252$&$0.0339948$\\
\hline
$10$&\qquad --&$0.0860492$&$0.0509321$&$0.0318852$&$0.0293065$\\
\hline
$12$&$-0.000986376$&$0.0332839$&$0.0469684$&$0.0274405$&$0.025918$\\
\hline
$14$&\qquad --&$-0.0156868$&$0.0420781$&$0.0238285$&$0.023602$\\
\hline
$16$&$-0.0158189$&$-0.0216262$&$0.0396338$&$0.0213232$&$0.0218065$\\
\hline
$18$&\qquad --&$-0.0308959$&$0.036239$&$0.0190955$&$0.0204342$\\
\hline
$20$&\qquad --&$-0.0253059$&$0.0339235$&$0.0174832$&$0.0193387$\\
\hline
\end{tabular}
\renewcommand{\arraystretch}{1}
\end{center}
\end{table}

\begin{figure}[htbp]
\begin{center}
\includegraphics[width=16cm]{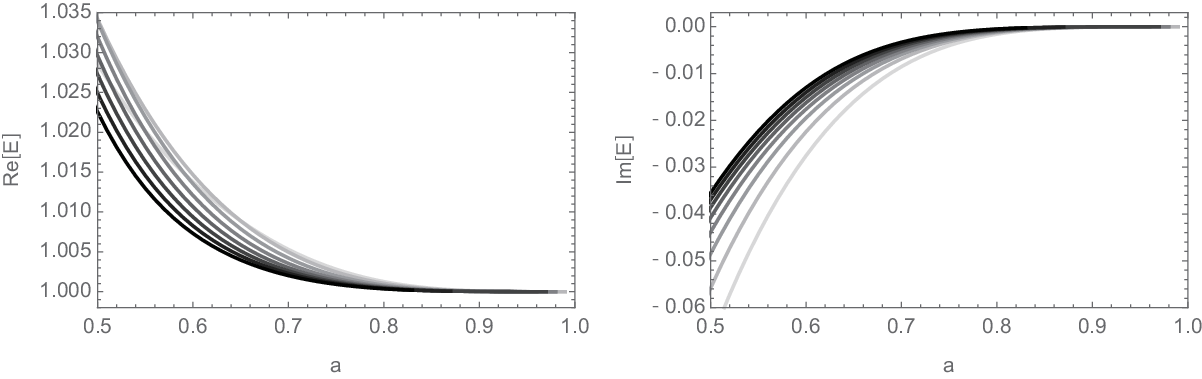}
\caption{Plots of the energy $E$ from Eq.~(\ref{eq:E-E0def}) of the ``double brane'' solution, $\Psi_{a}^{\rm D}$, truncated up to level $L=6,8,\dots,20$.
The left and right figures show the real and imaginary parts of $E$, respectively.
A darker line corresponds to a higher level.
These are a magnified version of Fig.~\ref{fig:ED} for $0.5\le a\le 1$.
\label{fig:EDa0p5to1}
}
\end{center}
\end{figure}

\begin{figure}[htbp]
\begin{center}
\includegraphics[width=16cm]{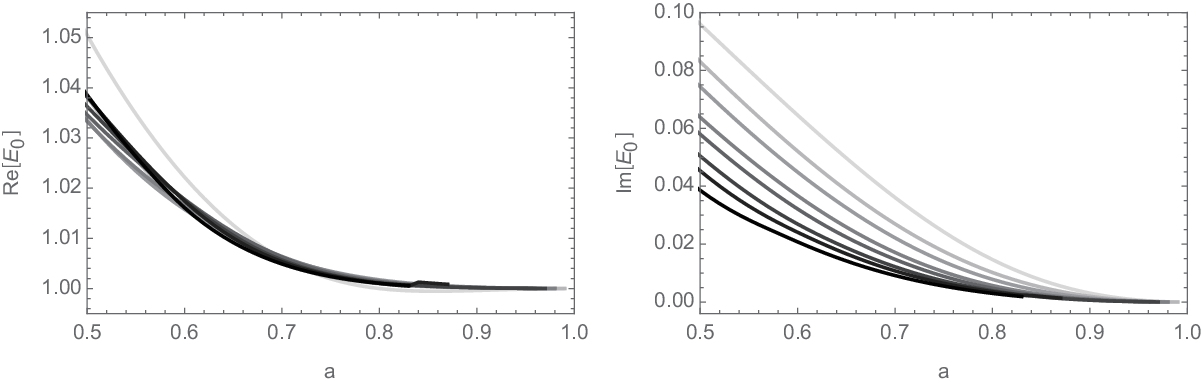}
\caption{Plots of the gauge-invariant observable $E_{0}$ from Eq.~(\ref{eq:E-E0def}) of the ``double brane'' solution, $\Psi_{a}^{\rm D}$, truncated up to level $L=6,8,\dots,20$.
The left and right figures show the real and imaginary parts of $E_{0}$, respectively.
A darker line corresponds to a higher level.
These are a magnified version of Fig.~\ref{fig:E0D} for $0.5\le a\le 1$.
\label{fig:E0Da0p5to1}
}
\end{center}
\end{figure}

The solution for the ``double brane,'' $\Psi_{a}^{\rm D}$, gives almost $1$ for $E$ and $E_{0}$ in $0.5<a<1$ as in Figs.~\ref{fig:ED} and \ref{fig:E0D}.
We demonstrate plots of $E[\Psi_{a}^{\rm D}]$ and $E_{0}[\Psi_{a}^{\rm D}]$ near $1$ for $0.5\le a\le 1$ in Figs.~\ref{fig:EDa0p5to1} and \ref{fig:E0Da0p5to1}, respectively.

\subsection{The validity of solutions in $a$-gauge
\label{sec:validityagauge}
}

\begin{figure}[htbp]
\begin{center}
\includegraphics[width=16cm]{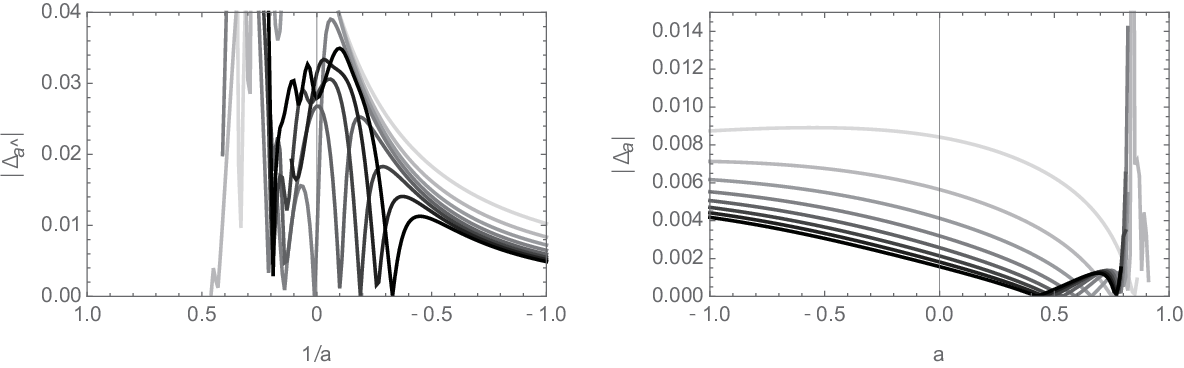}
\caption{The absolute value of the lowest level of the out-of-$a$-gauge equations, Eqs.~(\ref{eq:outofgaugehat}) or (\ref{eq:outofgauge}), for the tachyon vacuum solutions
$|\Delta_{\hat a}[\varPsi_{\hat a}^{\rm T}]|$ (left) and $|\Delta_a[\Psi_{a}^{\rm T}]|$ (right) for the truncation level $L=6,8,\dots,20$.
A darker line corresponds to a higher level.
\label{fig:DeltaaT}
}
\end{center}
\end{figure}

\begin{figure}[htbp]
\begin{center}
\includegraphics[width=16cm]{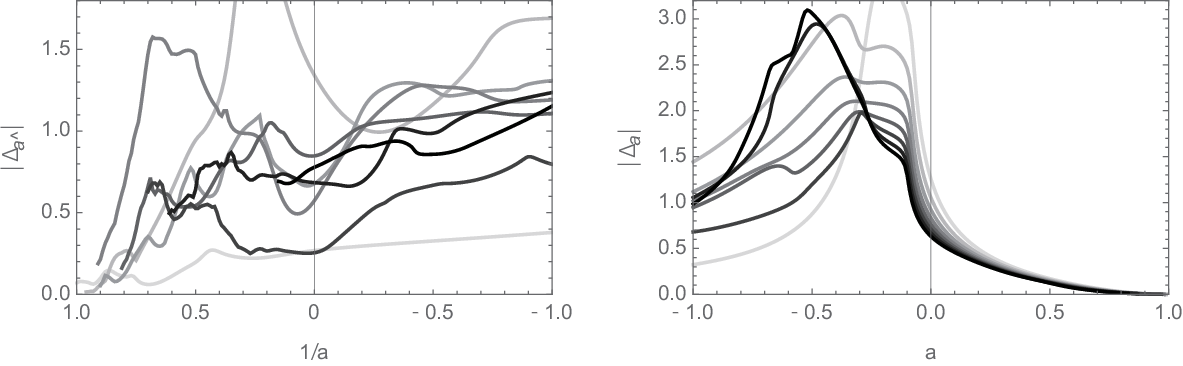}
\caption{The absolute value of the lowest level of the out-of-$a$-gauge equations, Eqs.~(\ref{eq:outofgaugehat}) or (\ref{eq:outofgauge}), for the ``double brane'' solutions
$|\Delta_{\hat a}[\varPsi_{\hat a}^{\rm D}]|$ (left) and $|\Delta_a[\Psi_{a}^{\rm D}]|$ (right) for the truncation level $L=6,8,\dots,20$.
A darker line corresponds to a higher level.
\label{fig:DeltaaD}
}
\end{center}
\end{figure}

\begin{figure}[htbp]
\begin{center}
\includegraphics[width=16cm]{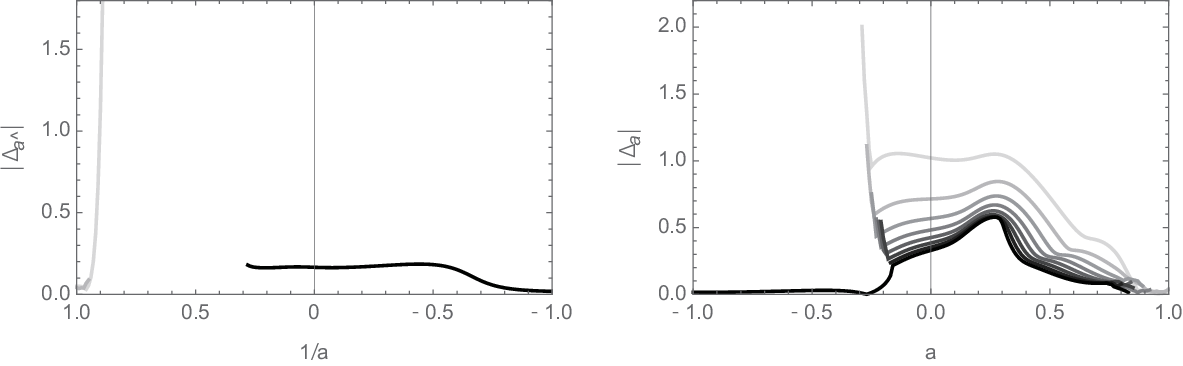}
\caption{The absolute value of the lowest level of the out-of-$a$-gauge equations, Eqs.~(\ref{eq:outofgaugehat}) or (\ref{eq:outofgauge}), for the ``ghost brane'' solutions
$|\Delta_{\hat a}[\varPsi_{\hat a}^{\rm G}]|$ (left) and $|\Delta_a[\Psi_{a}^{\rm G}]|$ (right) for the truncation level $L=6,8,\dots,20$.
A darker line corresponds to a higher level.
\label{fig:DeltaaG}
}
\end{center}
\end{figure}

As consistency checks of the solutions obtained in $a$-gauge, we evaluated the out-of-$a$-gauge equations in Eq.~(\ref{eq:outofgauge}) for $-1\le a\le 1$ or Eq.~(\ref{eq:outofgaugehat}) for $-1\le \hat{a}=1/a \le 1$.
 Here, we show plots of the absolute value of the lowest level of them, $|\Delta_{a}[\Psi_{a}]|$ or $|\Delta_{\hat a}[\varPsi_{\hat{a}}]|$, in Fig.~\ref{fig:DeltaaT} for the tachyon vacuum solution, Fig.~\ref{fig:DeltaaD} for the ``double brane'' solution, and Fig.~\ref{fig:DeltaaG} for the ``ghost brane'' solution.\\

\begin{figure}[htbp]
\begin{center}
\includegraphics[width=16cm]{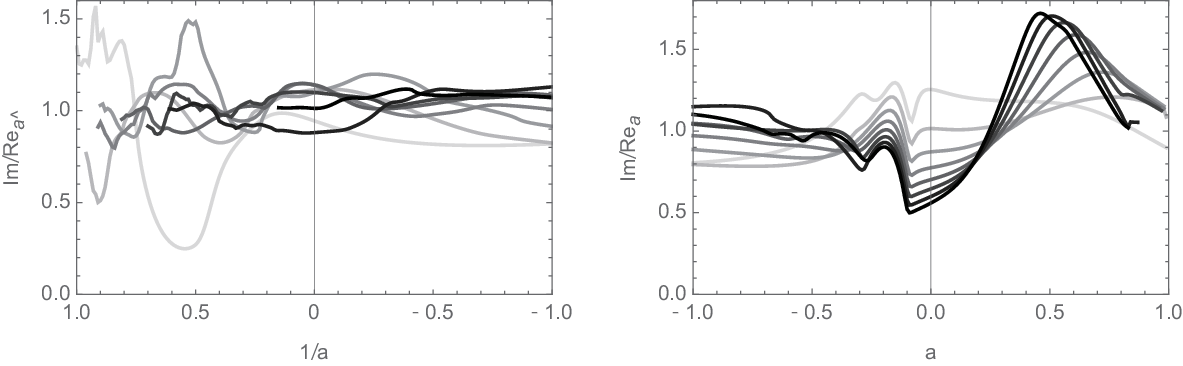}
\caption{${\rm Im}/{\rm Re}_{\hat a}[\varPsi_{\hat a}^{\rm D}]$ from Eq.~(\ref{eq:IoRahatdef}) (left) and ${\rm Im}/{\rm Re}_a[\Psi_{a}^{\rm D}]$ from Eq.~(\ref{eq:IoRadef}) (right) for the truncation level $L=6,8,\dots,20$.
A darker line corresponds to a higher level.
\label{fig:IoRD}
}
\end{center}
\end{figure}

\begin{figure}[htbp]
\begin{center}
\includegraphics[width=16cm]{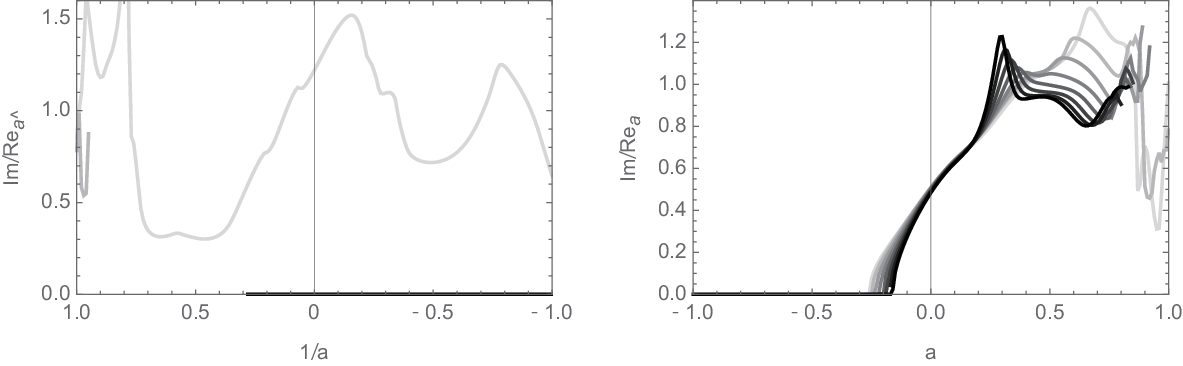}
\caption{${\rm Im}/{\rm Re}_{\hat a}[\varPsi_{\hat a}^{\rm G}]$ from Eq.~(\ref{eq:IoRahatdef}) (left) and ${\rm Im}/{\rm Re}_a[\Psi_{a}^{\rm G}]$ from Eq.~(\ref{eq:IoRadef}) (right) for the truncation level $L=6,8,\dots,20$.
A darker line corresponds to a higher level.
\label{fig:IoRG}
}
\end{center}
\end{figure}

Although the solutions for the tachyon vacuum in $a$-gauge, $\Psi_{a}^{\rm T}$ and $\varPsi_{\hat a}^{\rm T}$, staisfy the reality condition exactly,
the solutions for ``double brane'' and ``ghost brane'' in $a$-gauge are complex in general.
Hence, we show plots of the ratio of the imaginary part to the real one of the solutions obtained in $a$-gauge, ${\rm Im}/{\rm Re}_{a}[\Psi_{a}]$ from Eq.~(\ref{eq:IoRadef}) for $-1\le a\le 1$  and ${\rm Im}/{\rm Re}_{\hat a}[\varPsi_{a}]$ from Eq.~(\ref{eq:IoRahatdef}) for $-1\le \hat{a}=1/a\le 1$,
 in Fig.~\ref{fig:IoRD} for the ``double brane'' solution $\Psi_{a}^{\rm D}$, $\varPsi_{\hat a}^{\rm D}$ and Fig.~\ref{fig:IoRG} for the ``ghost brane'' solution $\Psi_{a}^{\rm G}$, $\varPsi_{\hat a}^{\rm G}$.

\bibliographystyle{utphys}
\bibliography{referencev3}

\end{document}